\documentclass[conference]{IEEEtran}
\usepackage{setspace}

\usepackage{amsmath}
\usepackage[dvips]{color}
\usepackage{comment}
\usepackage{todonotes}
\usepackage{epsf}
\usepackage{epsfig}
\usepackage{times}
\usepackage{epsfig}
\usepackage{graphicx}
\makeatletter
\def\widebreve{\mathpalette\wide@breve}
\def\wide@breve#1#2{\sbox\z@{$#1#2$}%
	\mathop{\vbox{\m@th\ialign{##\crcr
				\kern0.08em\brevefill#1{0.8\wd\z@}\crcr\noalign{\nointerlineskip}%
				$\hss#1#2\hss$\crcr}}}\limits}
\def\brevefill#1#2{$\m@th\sbox\tw@{$#1($}%
	\hss\resizebox{#2}{\wd\tw@}{\rotatebox[origin=c]{90}{\upshape(}}\hss$}
\makeatletter
\usepackage{mathrsfs}
\usepackage{amssymb}
\usepackage{pdfpages}
\usepackage{epstopdf}
\usepackage{authblk}
\usepackage{cleveref}
\usepackage{url}
\usepackage{dsfont}
\usepackage{lettrine} 
\usepackage{cite}
\usepackage{array}
\usepackage[ruled,lined,shortend,linesnumbered,fillcomment]{algorithm2e}
\usepackage{amsmath,epsfig,amssymb,algpseudocode,amsthm,cite,url}
\IEEEoverridecommandlockouts
\usepackage{subcaption}
\allowdisplaybreaks
\usepackage{csquotes}

\usepackage{verbatim}
\usepackage{xcolor}
\usepackage[english]{babel}
\usepackage{amsmath,amssymb}
\usepackage{makecell}

\usepackage[justification=centering]{caption}
\usepackage{verbatim}

\newtheorem{definition}{\bf Definition}

\def\@IEEEauthorblockAstyle{\smallfont\normalsize}
\newcommand*{\rom}[1]{\expandafter\@slowromancap\romannumeral #1@}
\begin{document}
\clearpage
\title{\fontsize{18.5pt}{23pt}\selectfont
	Resource Allocation for mmWave-NOMA Communication through Multiple Access Points Considering Human Blockages}
\author{Foad Barghikar, Foroogh S. Tabataba, and Mehdi Naderi Soorki
	\thanks{Foad Barghikar and Foroogh S. Tabataba are with the Department of Electrical and Computer Engineering, Isfahan University of Technology, Isfahan, Iran 84156-83111. Emails: f.barghikar@ec.iut.ac.ir and fstabataba@iut.ac.ir.}
	\thanks{Mehdi Naderi Soorki is with the Department of Electrical Engineering, Shahid Chamran University of Ahvaz, Ahvaz, Iran. Email: m.naderisoorki@scu.ac.ir.}}
\maketitle
\thispagestyle{empty}
\begin{abstract}
In this paper, a new framework for optimizing the resource allocation in a millimeter-wave-non-orthogonal multiple access (mmWave-NOMA) communication for crowded venues is proposed. MmWave communications suffer from severe blockage caused by obstacles such as the human body, especially in a dense region. Thus, a detailed method for modeling the blockage events in the in-venue scenarios is introduced. Also, several mmWave access points are considered in different locations. To maximize the network sum rate, the resource allocation problem is formulated as a mixed integer non-linear programming, which is NP-hard in general. Hence, a three-stage low-complex solution is proposed to solve the problem. At first, a user scheduling algorithm, i.e., modified worst connection swapping (MWCS), is proposed. Secondly, the antenna allocation problem is solved using the simulated annealing algorithm. Afterward, to maximize the network sum rate and guarantee the quality of service constraints, a non-convex power allocation optimization problem is solved by adopting the difference of convex programming approach. The simulation results show that, under the blockage effect, the proposed mmWave-NOMA scheme performs on average $23\%$ better than the conventional mmWave-orthogonal multiple access scheme. Moreover, the performance of proposed solution is $11.4\%$ lower than the optimal value while reducing complexity by $96\%$.
\end{abstract}
\section{Introduction}
Bandwidth shortage in the current wireless communications is one of the most critical barriers to meet the ever-increasing demand for high data rates in the 5G and beyond. Hence, a large amount of unoccupied bandwidth has recently been considered in the mmWave spectrum, i.e., the frequency range of 30 GHz to 300 GHz \cite{millimeterSurvey}. The regulatory agencies such as the federal communications commission (FCC) have authorized the use of various bands in the mmWave spectrum, including an unlicensed 60 GHz band with a 500 MHz bandwidth \cite{FCC}. However, in the mmWave communication, there are many challenges including high path loss, sensitivity to blockage caused even by user's bodies and objects, and lower path loss in line-of-sight (LoS) mode compared to non-LoS (NLoS) mode \cite{millimeterSurvey}. There are various techniques to alleviate these limitations such as dense access point (AP) deployment, i.e., small cells, applying beamforming, and the use of high-gain directional antennas \cite{millimeterSurvey}.

Besides using mmWave in the next-generation wireless networks, massive connectivity and high spectrum efficiency are new requirements that conventional orthogonal multiple access techniques cannot satisfy them \cite{ProtocolNOMA1, OMAvsNOMA}. In contrast to traditional orthogonal multiple access (OMA) schemes, NOMA in the power domain can provide a high spectrum efficiency \cite{ProtocolNOMA1, OMAvsNOMA,coexistence}. The main idea behind NOMA is sharing the same resources (e.g., frequency or time) by multiple users and separating them in the power domain using superposition coding (SC) at the transmitter and successive interference cancellation (SIC) at the receiver \cite{ProtocolNOMA1}. Given the above characteristics of the mmWave communication and NOMA, the use of the NOMA technique in the mmWave spectrum can bring many benefits \cite{OMAvsNOMA,coexistence}. However, applying the NOMA scheme in a multiple-AP mmWave network requires solving new complex problems such as optimal AP placement, beam steering, user grouping, antenna, and power allocation.
\subsection{Related works}
Multiple AP deployment is one of the well-known effective methods to increase the availability of the LoS mmWave links and optimize network coverage \cite{3AccessPointDeploymentMmWave,1AccessPointDeploymentMmWave}. The work in \cite{3AccessPointDeploymentMmWave} proposed a greedy algorithm to solve the AP deployment problem to maximize coverage. In \cite{1AccessPointDeploymentMmWave}, the authors have formulated a joint AP deployment and beam steering problem to minimize the number of required APs while the coverage constraints are met. However, the resource allocation problem has not been addressed in \cite{3AccessPointDeploymentMmWave} and \cite{1AccessPointDeploymentMmWave}. Recent works such as \cite{WCS,mmWaveCellular,multi-AP-60GHz} and \cite{Add-multicell-mmWave} have investigated the resource allocation problem and the user association for the multi-cell mmWave networks. Due to dense deployment in the mmWave cellular networks, the co-channel interference management is very challenging. In \cite{WCS}, the authors have introduced a load balancing user association scheme for mmWave cellular networks by defining an optimization problem to maximize the sum rate, and then they have proposed a heuristic algorithm to solve it. They have shown that the user association significantly affects network interference and users' instantaneous rates. In \cite{mmWaveCellular}, the authors have investigated a mmWave dense femtocell network. First, they have proposed a clustering scheme based on the LoS connectivity for co-channel interference management and then have applied the power and sub-channel allocation separately for users and femto APs in each cluster. In \cite{multi-AP-60GHz}, the user-AP association problem in a 60 GHz WLAN has been investigated with the goal of maximizing user rate. In \cite{Add-multicell-mmWave}, the authors have formulated the joint beam and power allocation in a multi-cell mmWave network as a mixed integer non-linear programming (MINLP) problem to maximize the network sum rate. It should be noted that none of the works in \cite{3AccessPointDeploymentMmWave,1AccessPointDeploymentMmWave,WCS,mmWaveCellular,multi-AP-60GHz,Add-multicell-mmWave} have applied the NOMA technique, and except in \cite{Add-multicell-mmWave} where the blockage effects have not been considered, the rest of these works in \cite{3AccessPointDeploymentMmWave,1AccessPointDeploymentMmWave,WCS,mmWaveCellular,multi-AP-60GHz} have considered a simple probabilistic model for the availability of LoS mmWave links.

Recent studies have used the NOMA scheme in a single-cell mmWave communication \cite{RandomBeamformingMmwave-NOMA, BeamwidthControl(EE)mmWave-NOMA, multi-beamNOMAmmWave, CorrelationGrouping2, JointTX-RX, Stackelberg}. When NOMA is applied to such a system, the issue of user grouping is raised. In this regard, in \cite{RandomBeamformingMmwave-NOMA, BeamwidthControl(EE)mmWave-NOMA, multi-beamNOMAmmWave, CorrelationGrouping2, JointTX-RX, Stackelberg}, the resource allocation, and user grouping methods for the single-cell mmWave-NOMA communications have been studied. In \cite{RandomBeamformingMmwave-NOMA}, the authors have used the random beamforming technique to steer beam toward users within each NOMA group. However, due to the narrow beams in mmWave communication, the random beamforming method cannot exploit all the potential of NOMA. In order to fix this problem, the authors in \cite{BeamwidthControl(EE)mmWave-NOMA} have proposed a beamwidth control method being appropriate for mmWave-NOMA communication and investigated its performance in terms of energy efficiency. The authors in \cite{multi-beamNOMAmmWave} have introduced the beam splitting technique for grouping the users located in different directions and then compared its performance with other existing methods in terms of sum rate. In \cite{CorrelationGrouping2}, the authors have firstly used a user grouping method based on channel correlation, and then they solved an analog beamforming problem with a boundary-compressed particle swarm algorithm to direct the beam to each user. In \cite{JointTX-RX}, the authors have introduced a joint transmitter-receiver beamforming and power allocation design for a mmWave-NOMA communication based on pure analog beamforming. In \cite{Stackelberg}, the authors have adopted a clustering approach for NOMA implementation using hybrid beamforming. It should be noted that the blockage effect is not considered in \cite{RandomBeamformingMmwave-NOMA}, \cite{multi-beamNOMAmmWave, CorrelationGrouping2, JointTX-RX}, and \cite{Stackelberg}. In \cite{BeamwidthControl(EE)mmWave-NOMA}, the blockage effect is modeled by a distance-dependent probabilistic model; however, the authors have not provided any solution to reduce this effect.

Although a multi-AP structure combined with the NOMA technique highly increases the performance of mmWave communication, the resource allocation problem is inherently complex in this network. This complexity increases if the multi-AP mmWave-NOMA network is deployed to cover a dense indoor region such as a lecture hall. In fact, due to the close deployment of APs and the dense arrangement of the users' seats, the inter-cell interference is no longer negligible. The resource allocation problem in the multi-cell mmWave-NOMA networks has been investigated in the recent works such as \cite{Multi-cellmmWaveNOMA,MultiCellmmWave-AngleDomainNOMA,Add-mulicell-mmWaveNOMA1,Add-mulicell-mmWaveNOMA2}. In \cite{Multi-cellmmWaveNOMA}, the authors have obtained a closed-form expression for the outage probability of multi-cell mmWave-NOMA networks and have taken a random user scheduling policy. In \cite{MultiCellmmWave-AngleDomainNOMA}, the authors have proposed an angle-domain NOMA scheme for the multi-cell mmWave networks without considering the blockage effects, in which two users located along an angle-domain beam are grouped together to implement NOMA technique. However, for greater spectrum efficiency and grouping of users in different directions, we use the beam splitting technique and antenna allocation in our model. The authors in \cite{Add-mulicell-mmWaveNOMA1} have solved the user association, sub-channel and power allocation issues for a mmWave-NOMA heterogeneous network (HetNet). In \cite{Add-mulicell-mmWaveNOMA2}, the authors have proposed a mmWave-NOMA HetNet along with machine-to-machine (M2M) communication. It should be noted that the blockage effects have not been considered in \cite{Add-mulicell-mmWaveNOMA1} and \cite{Add-mulicell-mmWaveNOMA2}, and both of these works have been focused on maximizing the energy efficiency of the network. 

In crowded venues, the random blockage effects from a large number of users' bodies should be considered in any user scheduling, antenna, and power allocation algorithms. Consequently, precise modeling of the user's body blockage affects the availability of mmWave links. Also, providing a resource allocation scheme using blockage model increases the validity of the design in the multi-AP mmWave-NOMA networks. Recent works have focused on modeling the blockage caused by large obstacles like buildings \cite{measurement_distance-dependent_blockage_model,variety_blockage_model}, and small objects such as human body \cite{foliage_body_blockage1,self-body-blocking,measurement_indoor_human_blockage,BlockageModeling2,complete_blockage_model}. In \cite{foliage_body_blockage1}, using ray tracing simulations a distance-based blockage probability function has been presented to model the blockage effects caused by human body and foliage. In \cite{self-body-blocking}, the self-body blockage has been modeled with a cone-blocking model, however, this model has not considered the effects of the nearby-user blockage. In \cite{measurement_indoor_human_blockage}, a stochastic human blockage model have been obtained by performing ray-launching simulations in a meeting room with a table in its center under realistic human movements. In \cite{BlockageModeling2}, the authors have introduced a method for modeling blockage events in a mmWave network consisting of smart phone and wearable devices such as smart watch, where the transmitter and receiver are usually adjacent to user's body. Therefore, the mathematical method for obtaining the nearby-user blockage in such scenarios are different from the in-venue scenarios in which the transmiter is far from the receiver and is installed on the ceiling. In \cite{complete_blockage_model}, the authors have presented an analytical model of the blockage effects caused by buildings, mobile blockers, and user's own body based on stochastic geometry. However, in an in-venue scenario, since there is no building and the location of users is fix, a deterministic analysis can be performed that is more accurate than their probabilistic analysis.
\subsection{Contributions}
In this paper, we consider a multi-AP mmWave-NOMA scheme and propose an appropriate framework for resource allocation in the downlink communication for dense indoor venues. The proposed approach allows the network to adjust the transmission beams of the mmWave AP, user scheduling, and power allocation in the presence of the stochastic user's body blockage over the mmWave links. In summary, our main contributions are as follows.
\begin{itemize}
	\item We precisely model the effect of human blockage in three-dimensional (3D) dense indoor scenarios to evaluate the performance of multi-AP mmWave-NOMA networks. In these 3D dense indoor environments, a large number of users are seated close to one another. Our 3D blockage modeling is applicable to a variety of real dense indoor venues, such as a lecture hall, sports stadium, and theater.
	\item We formulate the joint user scheduling, antenna, and power allocation in mmWave-NOMA networks as an optimization problem to maximize the users' sum rate under stochastic blockage effects and quality of service (QoS) constraints. Then, we propose a three-stage solution with low complexity to solve this optimization problem. In the first stage, the user scheduling is formulated as an optimization problem. Considering the channel correlation and channel gain difference, we develop a low complexity algorithmic solution, i.e., a modified worst connection swapping technique to assign users to mmWave APs and group them for NOMA implementation (i.e., user scheduling design). In the second stage, we apply a meta-heuristic algorithm, named simulated annealing (SA), to split the beams among users' NOMA group. In the third stage, given the obtained user scheduling and antenna allocation strategy in the previous stages, the power allocation problem is formulated as a non-convex optimization problem to maximize the system sum rate and guarantee the QoS constraints.
	\item We transform the formulated problem as a canonical form and solve the new canonical form power allocation problem based on the difference of convex (DC) programming approach. Simulation results show that the spectral efficiency of the proposed mmWave-NOMA system under the blockage effect is on average $23\%$ and $9.2\%$ higher than the corresponding OMA system with and without blockage, respectively. In addition, the performance of our proposed method is on average $11.4\%$ lower than the optimal solution, while reducing the computational complexity by about $96\%$.
\end{itemize}

The rest of the paper is organized as follows. In Section \ref{Sec:Sys-Model}, the system model is presented. The resource allocation problem for mmWave-NOMA communication is formulated in Section \ref{Sec:problem}. Section \ref{Sec:Reso_allo_Design} offers the three-stage low-complex solution for the formulated resource allocation problem. In Section \ref{Sec:Sim_results}, simulation results demonstrate the performance of the proposed solution, and a conclusion is presented in Section \ref{Sec:Conclusion}.
\section{System Model}\label{Sec:Sys-Model}
\subsection{System model}
We consider a downlink mmWave-NOMA communication in a dense indoor scenario composed of one sub-6GHz access point, a set $\mathcal{B}$ of $B$ mmWave access points (mm-APs), and a set $\mathcal{K}$ of $K$ mobile devices (MDs). Each mm-AP and MD are equipped with an uniform linear antenna (ULA) array with $M_\text{AP}$ and $M_\text{MD}$ antenna elements, respectively. A fully connected hybrid structure is used in the mm-APs and MDs. Due to the power limitation, each MD has only one radio frequency (RF) chain, which is connected to all of its $M_\text{MD}$ antenna elements via $M_\text{MD}$ phase shifters (PSs). Each mm-AP is equipped with a set $\mathcal{N}$ of $N$ RF chains where each RF chain $n$ is connected to all $M_\text{AP}$ antenna elements through $M_\text{AP}$ PSs. In this network, the sub-6GHz AP acts as a central control unit that sends the required control signals \cite{dual-connectivity}. Since mmWave links are susceptible to the blockage and therefore are not reliable for exchanging the control signals, MDs use the sub-6GHz links to exchange control signals between themselves and sub-6GHz AP. In our model, mm-APs connect to the central controller for sending channel state information (CSI) and receiving control signals via the backhaul link. To reduce the probability of MDs being exposed to LoS link blockage, we deploy several mm-APs in different locations.

\begin{figure}[t!]
	\begin{center}
		\includegraphics[height=4.9cm,width=8.7cm]{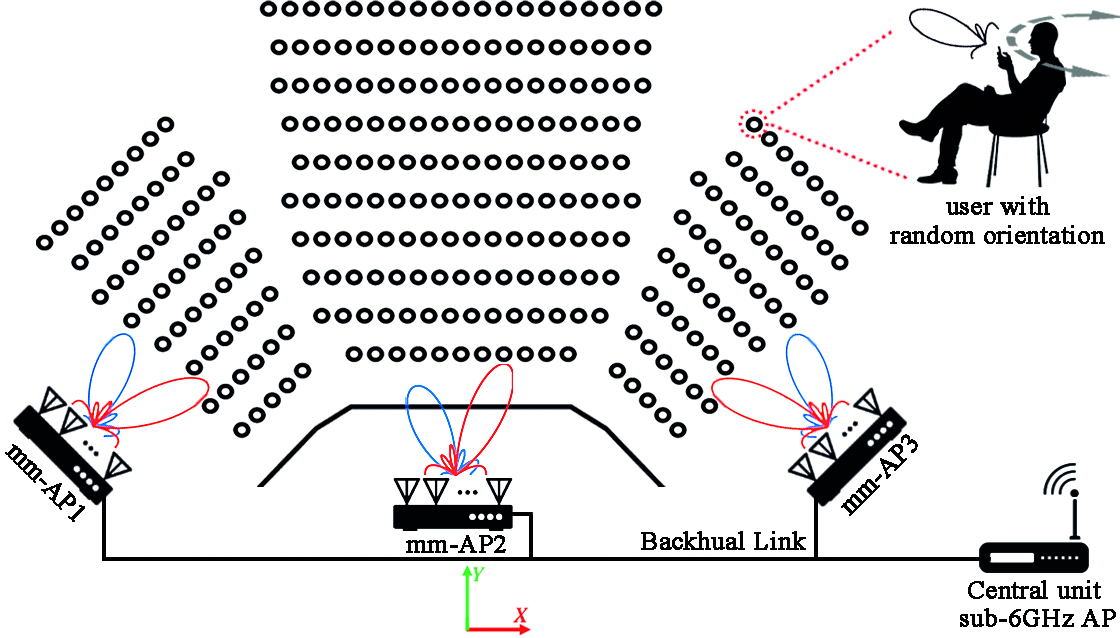}
	\end{center}
	\vspace{-0.7em}
	\caption{\footnotesize The system model for a mmWave-NOMA network in a venue.}
	\label{fig1}
\end{figure}
In our model, we focus on in-venue scenarios in which a high number of active users sit on a fixed and densely-deployment seating chart such as in a sports stadium, a lecture hall, a concert venue, or a theater. Thus, based on the arrangement and direction of the seats inside a given region, we assume that $K$ users sit on the seats and the location of each user $k \in \mathcal{K}$ is given by $(X_k,Y_k,H_k)$ in 3D Cartesian coordinates, where $H_k$ is equal with the height of user $k$ in sitting position plus the height of the platform on which the user $k$'s seat is located. Fig. \ref{fig1} shows an illustrative example of our proposed model. Due to the random changes in the orientation of the users within the horizon plane, the azimuthal angle of a given user $k$ is assumed to be a random variable, denoted by $\widetilde{\Psi}_k$ (See Fig. \ref{fig2}), with a given probability distribution function $\Pr(\widetilde{\Psi}_k)$, where $\widetilde{\Psi}_k \in [0,2\pi]$ that can practically be obtained for a given venue \cite{OrientationMeasurementLiFi}.

In our system, all users within a NOMA group are served by one RF chain. Thus, the users grouping is done to form NOMA groups in each mm-AP. We define a binary decision variable $c_{kbn}$ where $c_{kbn}=1$ if user $k$ is assigned to the mm-AP $b$ and RF chain $n$, otherwise $c_{kbn}=0$. Note that, to reduce the time delay and error propagation of SIC decoding, we assume that the maximum capacity of each NOMA group is two users, i.e., $\sum_{k=1}^{K}c_{kbn}\leq 2,\forall b \in \mathcal{B}$ and $\forall n \in \mathcal{N}$, \cite{BeamwidthControl(EE)mmWave-NOMA} and \cite{multi-beamNOMAmmWave}. In addition, we assume that each user can be allocated to at most one mm-AP and at most one RF chain of that mm-AP, i.e., $\sum_{b=1}^{B}\sum_{n=1}^{N}c_{kbn}\leq 1,\forall k\in\mathcal{K}$. Following the beam splitting technique \cite{multi-beamNOMAmmWave}, the antenna allocation among users in a NOMA group is required. The beam spilitting technique will be described in sec. \ref{subsec:beam}. $M_{kbn}$ represents the number of antennas allocated to user $k$ associated with RF chain $n$ in mm-AP $b$ in such a way that $\sum_{k=1}^{K}c_{kbn}M_{kbn}\leq M_\text{AP}, \forall b \in \mathcal{B}$ and $\forall n \in \mathcal{N}$. Let $\mathbf{G}_b=[\mathbf{g}_{b1},\dots,\mathbf{g}_{bN}]\in \mathbb{C}^{N\times N}$ be the digital precoder used in mm-AP $b$ where $\mathbf{g}_{bn}$ with $\|\mathbf{g}_{bn}\|^2=1$ denotes digital precoder related to the NOMA group associated with RF chain $n$. $\mathbf{x}_b=[x_{b1},\dots,x_{bN}]^\text{T}\in\mathbb{C}^{N\times 1}$ shows a signal vector transmitted from mm-AP $b$ and $x_{bn}=\sum_{k=1}^{K}c_{kbn}\sqrt{p_{kbn}}s_k$ is the superimposed signal of the NOMA group associated with RF chain $n$, where $s_k\in \mathbb{C}$ denotes the modulated symbol for user $k$ and $p_{kbn}$ denotes the power allocated to user $k$ in mm-AP $b$ and RF chain $n$. In addition, we have the sum power constraint, $\sum_{k=1}^{K}\sum_{b=1}^{B}\sum_{n=1}^{N}c_{kbn}p_{kbn}\leq p_\text{total}$, where $p_\text{total}$ is the maximum power budget for all the mm-APs. Here, we focus on the overloaded scenarios, i.e., $BN\leq K$.
\subsection{Beam splitting technique}\label{subsec:beam}
The beam splitting technique is used to divide the beam of each RF chain in different directions \cite{multi-beamNOMAmmWave}. Thus, to group the users located in different directions, we use the beam splitting technique in our model. In this case, the total number of ULA antenna elements are divided between the users that form a group. Then, the users' superimposed signal are transmitted by one RF chain through all antenna elements of ULA. However, the phase shifters corresponding to the allocated antenna elements to each user are adjusted according to the angle of departure (AoD) corresponding to that user. In this case, analog beamforming vector used by $n$-th RF chain of mm-AP $b$, $\mathbf{w}_{bn}\in \mathbb{C}^{M_\text{AP}\times1}$, is given by \cite{multi-beamNOMAmmWave}
\begin{align}
&\mathbf{w}_{bn}=[\widetilde{\mathbf{w}}^\text{T}(c_{1bn},\theta_{1b0},M_{1bn}),\dots,\widetilde{\mathbf{w}}^\text{T}(c_{Kbn},\theta_{Kb0},M_{Kbn})]^\text{T},\notag\\
&\text{in which}\notag\\
&\widetilde{\mathbf{w}}(c_{kbn},\theta_{kb0},M_{kbn})=\notag\\
&\left\{\begin{array}{ll}
\emptyset,&c_{kbn}=0,\\
e^{j\sum_{d=1}^{k-1}c_{dbn}M_{dbn}\zeta(\theta_{db0})}\mathbf{w}(\theta_{kb0},M_{kbn}),&c_{kbn}=1,
\end{array}\right.
\end{align}
where, $\mathbf{w}(\theta_{kb0},M_{kbn})=\frac{1}{\sqrt{M_\text{AP}}}[1,\dots,e^{j(M_{kbn}-1)\zeta(\theta_{kb0})}]^\text{T}$ is the analog beamformer for the $M_{kbn}$ antenna sub-array assigned to user $k$ in mm-AP $b$ and RF chain $n$.
Here, $\theta_{kb0}$ is the AoD towards user $k$ at the mm-AP $b$ for the LoS component. Note that $\zeta(\theta)=\frac{2\pi d}{\lambda}\cos(\theta)$, in which $\lambda$ shows the wavelength of the carrier frequency, and $d=\frac{\lambda}{2}$ is the distance between two adjacent elements of the antenna.
\subsection{Blockage model}
In our model, LoS mmWave links are subjected to both self-body blockages from a given user as well as a blockage from nearby users. The self-body blockage resulting from the user's body on his or her device is dynamic because the user's orientation randomly changes. On the other hand, the blockage due to the walls and nearby seats in an in-venue region is static. Due to the use of ULA (located in the horizontal plane) in the transceiver and since ULA propagation pattern only changes on the horizontal plane and is constant on the elevation plane \cite{AntennaArray}, we consider 2D model here. In the following, we model the human body with a cylinder \cite{BlockageModeling2}.
\subsubsection{Self-body blockage} 
\begin{figure}[t]
	\begin{center}
		\includegraphics[height=3.2cm,width=8.7cm]{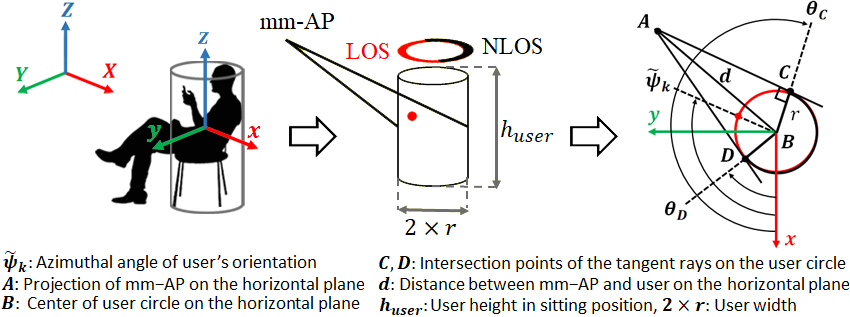}
	\end{center}
	\vspace{-0.7em}
	\caption{\footnotesize Self-body blockage modeling.}
	\label{fig2}
\end{figure}

To determine the self-body blockage for user $k$ relative to mm-AP $b$, we define $\mathcal{A}_{kb,1}$ as a set of azimuthal angles that the body of user $k$ does not block the LoS link. The complement of $\mathcal{A}_{kb,1}$ specifies the self-body blockage angles. According to Fig. \ref{fig2}, we can obtain angles $\theta_C$ and $\theta_D$ using the coordinates of the points $A$ and $B$ and the radius $r$. Consequently, $\mathcal{A}_{kb,1}=[\theta_D,\theta_C]$.
\subsubsection{Nearby-user blockage} We define set $\mathcal{A}_{kb,2}$ to describe a set of angles in which the LoS link between user $k$ and mm-AP $b$ is not blocked by the users sitting nearby of the user. The steps for obtaining this set are as follows.

\begin{figure}[t!]
	\begin{center}
		\includegraphics[height=6.2cm,width=5.1cm]{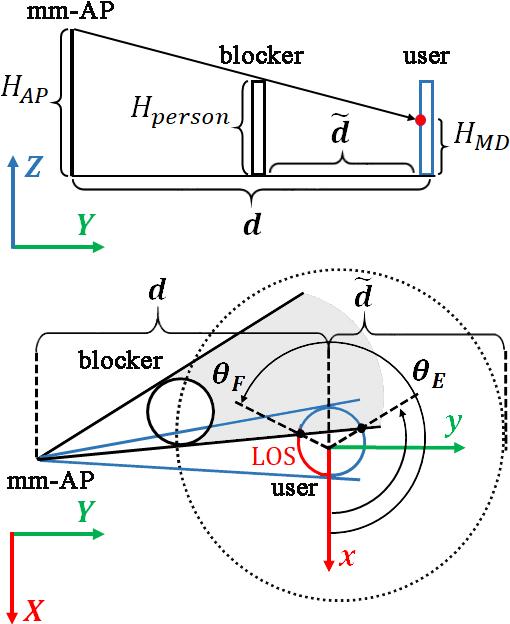}
	\end{center}
	\vspace{-0.7em}
	\caption{\footnotesize Nearby-user blockage modeling.}
	\label{fig3}
\end{figure}

Step 1: as it can be seen in Fig. \ref{fig3}, the effective distance $\widetilde{d}$ between user $k$ and a person with the height $H_\text{person}$ around it, is given by $\widetilde{d}=\frac{H_\text{person}-H_\text{MD}}{H_\text{AP}-H_\text{MD}}d$, where $H_\text{AP}$ and $H_\text{MD}$ denote the heights of mm-AP and MD, respectively; note that if users' seats are on a platform above the ground level, the height of the platform must also be added to $H_\text{person}$ and $H_\text{MD}$.\\
\indent Step 2: we draw a circle with center $(X_k,Y_k)$ and radius $\widetilde{d}$; then among the users within this circle, we only consider those who are taller than or equal to $H_\text{person}$ and their distance to mm-AP $b$ is less than $d$.\\
\indent Step 3: for users holding step 2 conditions, we draw the tangent lines to each of their circles and consider only those lines that pass through the user circle.\\
\indent Step 4: for the lines considered in step 3, we obtain the coordinates of the intersection of these lines with the user circle. We then calculate the angles corresponding to these points, such as $\theta_E$ and $\theta_F$ shown in Fig. \ref{fig3}.\\
\indent Step 5: users corresponding to the lines considered in step 3 can create a shadow for the user. Consequently, complement of $\mathcal{A}_{kb,2}$ is equal to the set of azimuthal angles that characterize the shaded areas. For example, for Fig. \ref{fig3}, $\mathcal{A}_{kb,2} = [\theta_F,2\pi]\bigcup[0,\theta_E]$.\\
\indent After repeating the above steps and defining set $\mathcal{A}_{kb}\triangleq\mathcal{A}_{kb,1}\bigcap\mathcal{A}_{kb,2}$, we can comment on the presence or absence LoS link between user $k$ and mm-AP $b$. For instance, the angles corresponding to the red part of the user circle in Fig. \ref{fig3} are equivalent to $\mathcal{A}_{kb}$ i.e., LOS angles. Considering the blockage effect, we define a binary variable $e_{kb}$ where $e_{kb}=0$ if the LoS link between user $k$ and mm-AP $b$ is blocked, otherwise $e_{kb}=1$.
\subsection{Channel model}
The channel matrix between the mm-AP $b$ and user $k$ can be represented as \cite{ChannelModel1}
\begin{align}
\mathbf{H}_{kb}=&e_{kb}\underbrace{\sqrt{\rho_{kb0}}\alpha_{kb0}\,\mathbf{a}_\text{MD}(\phi_{kb0})\mathbf{a}_\text{AP}^\text{H}(\theta_{kb0})}_{\text{LoS component}}\notag\\
&+\underbrace{\sum_{l=1}^{L}\sqrt{\rho_{kbl}}\alpha_{kbl}\,\mathbf{a}_\text{MD}(\phi_{kbl})\mathbf{a}_\text{AP}^\text{H}(\theta_{kbl})}_{\text{Scattering components}},
\end{align}
where $\rho_{kbl}$ and $\alpha_{kbl}\sim \mathcal{CN}(0,1)$ denote the average path loss and complex gain of the $l$-th path between user $k$ and mm-AP $b$, respectively. Here, $l=0$ represents the LoS path and $l\in\{1,\dots,L\}$ represents the $l$-th NLoS path. $\rho_{kbl}=(\frac{c}{4\pi f_c})^2d_{kb}^{-\gamma_l}$ in which $c=3\times10^8
$ m/s, $f_c$ represents the carrier frequency, $d_{kb}$ shows the distance of between user $k$ and mm-AP $b$, and $\gamma_l$ denotes the path loss exponent of $l$-th path.
$\mathbf{a}_\text{AP}(\theta_{kbl})=[1,e^{j\zeta(\theta_{kbl})},\dots,e^{j(M_\text{AP}-1)\zeta(\theta_{kbl})}]^\text{T}\in\mathbb{C}^{M_\text{AP}\times 1}$ denotes the array response vector for the  $l$-th path with AoD $\theta_{kbl}$ and $\mathbf{a}_\text{MD}(\phi_{kbl})=[1,e^{j\zeta(\phi_{kbl})},\dots,e^{j(M_\text{MD}-1)\zeta(\phi_{kbl})}]^\text{T}\in \mathbb{C}^{M_\text{MD}\times 1}$ shows the array response vector for the  $l$-th path with angle of arrival (AoA) $\phi_{kbl}$ from mm-AP $b$ at user $k$. We call the channel between mm-AP and MD RF chains as the effective channel. Consequently, the effective channel for the user $k$ allocated to $n$-th RF chain of mm-AP $b$  is defined as follows
\begin{equation}
\widetilde{h}_{kbn}=\mathbf{v}^\text{H}_k\mathbf{H}_{kb}\mathbf{w}_{bn},
\end{equation}
where $\mathbf{v}_k\in \mathbb{C}^{M_\text{MD}\times1}$ is the analog beamforming vector at user $k$ where $\mathbf{v}_k=\emptyset$ if $c_{kbn=0}$, otherwise $\mathbf{v}_k=\frac{1}{\sqrt{M_{MD}}}[1,\dots,e^{j(M_\text{MD}-1)\zeta(\phi_{kb0})}]^\text{T}$. Likewise, the effective channel vector of user $k$ is defined as $\mathbf{\widetilde{h}}_{kb}=[\widetilde{h}_{kb1},\dots,\widetilde{h}_{kbN}]^T\in \mathbb{C}^{N\times 1}$ and given the set $\mathcal{K}_b$ as the collection of $K_b$ users served by mm-AP $b$, the effective channel matrix for this mm-AP is defined as $\mathbf{\widetilde{H}}_b=[\mathbf{\widetilde{h}}_{1b},\dots,\mathbf{\widetilde{h}}_{K_bb}]\in \mathbb{C}^{N\times K_b}$. In our model, we modify the three-step low-complexity single-cell channel estimation algorithm presented in \cite{ChannelEstimation} to be applied in a multi-AP mmWave network. In step 1, MDs transmit unique frequency tones via one of the omnidirectional antennas in the antenna array to the mm-APs. Then, the mm-APs send estimated $\hat{\theta}_{kb0}$ and $\hat{\alpha}_{kb0}$ (LoS CSI) via backhaul link to the central controller. In step 2, each mm-AP during assigned time slot sends unique frequency tones to all the users using analog beamforming adjusted with $\hat{\theta}_{kb0}$. Then, MDs use the estimated $\hat{\phi}_{kb0}$ to form analog beamforming vector on the user side and transmit $\hat{\phi}_{kb0}$ over the sub-6GHz link to the central controller for next decisions. In step 3, 
the effective channel is estimated after user scheduling and antenna allocation in the central controller and sent to the users and mm-APs. In addition, digital beamforming and power allocation are done by exploiting the effective channel. Following time division duplex  and channel reciprocity, in our model,  the estimated effective channel in uplink can be used for designing the digital precoder in the downlink.
\subsection{SIC decoding order}
Without loss of generality, we assume that the users are indexed in the descending order of their LoS gains \cite{BeamwidthControl(EE)mmWave-NOMA,multi-beamNOMAmmWave}, i.e.,
\begin{equation}\label{eq4}
\rho_{1b_10}|\alpha_{1b_10}|^2\geq\rho_{2b_20}|\alpha_{2b_20}|^2\geq,\dots,\geq\rho_{Kb_K0}|\alpha_{Kb_K0}|^2,
\end{equation}
where $b_k\in\mathcal{B}$ denotes the mm-AP index assigned to user $k$. Note that if the LoS path is blocked, we will replace the NLoS path with the highest gain with the LoS path. We consider a fixed SIC decoding order according to the LoS path gain order $1,\dots,K$ to have a tractable resource allocation scheme, and to decrease the system overhead.

In our model the received signal vector at the users, $\mathbf{y}=[y_1,\dots,y_K]^\text{T}\in\mathbb{C}^{K\times1}$, is given by
\begin{equation}
\mathbf{y}=\widetilde{\mathbf{H}}^\text{H}\mathbf{G}\mathbf{x}+\boldsymbol{\nu},
\end{equation}
where $\widetilde{\mathbf{H}}^\text{H}\triangleq\left[\widetilde{h}_{kbn}^{*}\right]\in\mathbb{C}^{K\times BN}$, so that the vectors $\widetilde{\mathbf{h}}_{k1}^\text{H}$ to $\widetilde{\mathbf{h}}_{kB}^\text{H}$ form the $k$-th row of this matrix, $\mathbf{G}\in\mathbb{C}^{BN\times BN}$ is a block diagonal matrix such that the matrices $\mathbf{G}_1$ to $\mathbf{G}_B$ are its main-diagonal blocks, $\mathbf{x}\triangleq[\mathbf{x}_1^\text{T},\dots,\mathbf{x}_B^\text{T}]^\text{T}\in\mathbb{C}^{BN\times 1}$ and $\boldsymbol{\nu}=[\nu_1,\dots,\nu_K]^\text{T}\in\mathbb{C}^{K\times1}$ is the noise vector. After some simplification, the received signal at user $k$, $y_k$, is given by
\begin{align}\label{eq6}
y_{k}=&\underbrace{\mathbf{\widetilde{h}}_{kb}^\text{H}\mathbf{g}_{bn}\sqrt{p_{kbn}}x_k}_{\text{Desired signal}}+\underbrace{\mathbf{\widetilde{h}}_{kb}^\text{H}\mathbf{g}_{bn}\sum_{j\in\mathcal{K},j\neq k}c_{jbn}\sqrt{p_{jbn}}x_j}_{\text{Intra-group interference}}\notag\\
&+\underbrace{\mathbf{\widetilde{h}}_{kb}^\text{H}\sum_{n^\prime\in\mathcal{N},n^\prime\neq n}\mathbf{g}_{bn^\prime}\sum_{j\in\mathcal{K}}c_{jbn^\prime}\sqrt{p_{jbn^\prime}}x_j}_{\text{Inter-group interference}}\notag\\
&+\underbrace{\sum_{b^\prime\in\mathcal{B},b^\prime\neq b}\mathbf{\widetilde{h}}_{kb^\prime}^\text{H}\mathbf{G}_{b^\prime}\mathbf{x}_{b^\prime}}_{\text{Inter-AP interference}}+\underbrace{\nu_k}_{\text{Noise}},
\end{align}
where $\nu_k\sim\mathcal{CN}(0,\sigma^2)$ denotes additive white Gaussian noise at user $k$ with the power of $\sigma^2$. In \eqref{eq6}, the first term represents the desired signal of user $k$, the second term denotes intra-group interference caused by the other users within the NOMA ‌group associated with $n$-th RF chain of mm-AP $b$, the third term called inter-group interference originates from all the other RF chains of mm-AP $b$, and the fourth term expresses the inter-AP interference induced by other mm-APs. Per traditional downlink NOMA protocol, the individual rate for user $k$ received from $n$-th RF chain of mm-AP $b$ is given by
\begin{align}
R_{k\rightarrow k}^{\,bn}&=\log_2\left(1+\frac{c_{kbn}p_{kbn}|\mathbf{\widetilde{h}}_{kb}^\text{H}\mathbf{g}_{bn}|^2}{\text{\rom{1}}_{k\rightarrow k}^{bn}+\text{\rom{2}}_{k}^{bn}+\text{\rom{3}}_{k}^{b}+\sigma^2}\right),\;\;
\text{where}\label{eq7}\\
\text{\rom{1}}_{k\rightarrow k}^{bn}&=|\mathbf{\widetilde{h}}_{kb}^\text{H}\mathbf{g}_{bn}|^2\sum_{j=1}^{k-1}c_{jbn}p_{jbn},\\
\text{\rom{2}}_{k}^{bn}&=\sum_{n^\prime\in\mathcal{N},n^\prime\neq n}|\mathbf{\widetilde{h}}_{kb}^\text{H}\mathbf{g}_{bn^\prime}|^2\sum_{j\in\mathcal{K}}c_{jbn^\prime}p_{jbn^\prime},\;\text{and}\\ \text{\rom{3}}_{k}^{b}&=\sum_{b^\prime\in\mathcal{B},b^\prime\neq b}\sum_{n\in\mathcal{N}}|\mathbf{\widetilde{h}}_{kb^\prime}^\text{H}\mathbf{g}_{b^\prime n}|^2\sum_{j\in\mathcal{K}}c_{jb^\prime n}p_{jb^\prime n},
\end{align}
are intra-group interference, inter-group interference, and inter-AP interference, respectively. Now, the individual data rate of user $k$ is computed as $R_k=\sum_{b\in\mathcal{B}}\sum_{n\in\mathcal{N}}R_{k\rightarrow k}^{bn}, \forall k\in\mathcal{K}$. If both users $k$ and $i$ are assigned to $n$-th RF chain of mm-AP $b$, $\forall k<i$, user $k$ first decodes the signal of user $i$ with the corresponding achievable rate given by
\begin{equation}
R_{k\rightarrow i}^{\,bn}=\log_2\left(1+\frac{c_{ibn}p_{ibn}|\mathbf{\widetilde{h}}_{kb}^\text{H}\mathbf{g}_{bn}|^2}{\text{\rom{1}}_{k\rightarrow i}^{bn}+\text{\rom{2}}_{k}^{bn}+\text{\rom{3}}_{k}^{b}+\sigma^2}\right).
\end{equation}

We define $\text{\rom{1}}_{k\rightarrow i}^{bn}\triangleq|\mathbf{\widetilde{h}}_{kb}^\text{H}\mathbf{g}_{bn}|^2\sum_{j=1}^{i-1}c_{jbn}p_{jbn}$ as the intra-group interference from the users of RF chain $n$ to user $k$ when decoding the signal of user $i$. To ensure successful SIC, the following rate condition must be guaranteed:
\begin{equation}
R_{i\rightarrow i}^{bn}\leq R_{k\rightarrow i}^{bn},\qquad\forall k<i.
\end{equation}

Consequently, the achievable sum rate for this system can be calculated by $ R_{\text{sum}}=\sum_{k\in\mathcal{K}}\sum_{b\in\mathcal{B}}$\\$\sum_{n\in\mathcal{N}}R_{k\rightarrow k}^{bn}.
$
\section{Problem Formulation}\label{Sec:problem}
Due to self-body and user-body blockages in highly dense venues, designing an optimal method of user scheduling, power allocation, and antenna allocation to maximize the sum rate of mmWave-NOMA network subject to the QoS constraints and the total power constraint is essentially required. Thus, we formulate the problem of mmWave-NOMA for an in-venue region as follows.
\begin{subequations}\label{eq13}
	\begin{align}
	&\underset{\left\{\substack{c_{kbn},p_{kbn},M_{kbn},\\k\in\mathcal{K},b\in\mathcal{B},n\in\mathcal{N}}\right\}}{\max}\,\sum_{k\in\mathcal{K}}\sum_{b\in\mathcal{B}}\sum_{n\in\mathcal{N}}R_{k\rightarrow k}^{bn},\\
	\text{s.t.}\,&\sum_{k=1}^{K}c_{kbn}\leq 2,\forall b\in\mathcal{B},\forall n\in\mathcal{N},\label{eq13c}\\
	&\sum_{b=1}^{B}\sum_{n=1}^{N}c_{kbn}\leq 1,\forall k\in\mathcal{K},\label{eq13d}\\
	&\sum_{k=1}^{K}c_{kbn}M_{kbn}\leq M_{\text{AP}},\forall b\in\mathcal{B},\forall n\in\mathcal{N},\label{eq13e}\\
	&c_{kbn}M_{\min}\leq c_{kbn}M_{kbn},\forall k\in\mathcal{K},\forall b\in\mathcal{B},\forall n\in\mathcal{N},\label{eq13f}\\
	&0\leq p_{kbn},\forall k\in\mathcal{K},\forall b\in\mathcal{B},\forall n\in\mathcal{N},\label{eq13g}\\
	&\sum_{k=1}^{K}\sum_{b=1}^{B}\sum_{n=1}^{N}c_{kbn}p_{kbn}\leq p_\text{total},\label{eq13h}\\
	&c_{kbn}R_{i\rightarrow i}^{bn}\leq c_{kbn}R_{k\rightarrow i}^{bn},\forall k<i,\forall b\in\mathcal{B},\forall n\in\mathcal{N},\label{eq13i}\\
	&c_{kbn}R_{\min}\leq \sum_{b=1}^{B}\sum_{n=1}^{N}R_{k\rightarrow k}^{bn},\forall k\in\mathcal{K},\label{eq13j}\\
	&c_{kbn}\in\{0,1\},\forall\,k\in\mathcal{K},\forall b\in\mathcal{B},\forall n\in\mathcal{N},\label{eq13k}\\
	&M_{kbn}\in\mathbb{N}.\label{eq13l}
	\end{align}
\end{subequations}

Constraint \eqref{eq13c} shows that the maximum capacity of each NOMA group is two users. Constraint \eqref{eq13d}  indicates that each user can be assigned to at most one mm-AP and one RF chain. Constraint \eqref{eq13e} guarantees that the number of all allocated antenna elements on $n$-th RF chain of mm-AP $b$ cannot be larger than $M_{\text{AP}}$. Constraint \eqref{eq13f} is a lower bound on the number of antennas allocated to each user. Also, constraint \eqref{eq13f} prevents the formation of a beam with high sidelobe followed by high co-channel interference. Constraint \eqref{eq13h} is the total transmission power constraint in all of the mm-APs. Constraint \eqref{eq13i} guarantees successful SIC and constraint \eqref{eq13j} is the QoS constraint with the predefined threshold $R_{\min}$ as the minimum rate requirement of each user. Due to the existence of continuous and integer variables as well as nonlinear functions in the objective function and constraints of the problem \eqref{eq13}, this problem is MINLP and in general, NP-hard \cite{MINLP}. Consequently, directly solving problem \eqref{eq13} is challenging and intractable. To achieve an efficient solution with acceptable computational complexity, we propose a three-stage low-complex method, which is detailed in section \ref{Sec:Reso_allo_Design}. 
\section{Resource Allocation}\label{Sec:Reso_allo_Design}
We decompose problem \eqref{eq13} into three sub-problems and propose a three-stage sub-optimal solution with low complexity to solve it. In the first stage, we set the user scheduling variable, $c_{kbn}$, to maximize the system sum rate, $R_{\text{sum},1}$, based on LoS CSI (channel without scattering components), users' channel correlation, users' channel gain difference, and modified worst connection swapping (MWCS) algorithm. At this stage, the power allocation and antenna allocation are considered to be uniform, for example if $c_{kbn}=1$ then $p_{kbn}={p_\text{total}}/{K}$ also if $c_{jbn}=c_{ibn}=1$, then $M_{jbn}=M_{ibn}={M_\text{AP}}/{2}$. We also assume that the digital precoder is unavailable, i.e., $\mathbf{G}_b=\mathbf{I}_N,\forall b\in\mathcal{B}$. In the second stage, and for NOMA groups with more than one member, the antenna allocation is performed using a meta-heuristic algorithm called simulated annealing to maximize the system sum rate in this stage, $R_{\text{sum},2}$, \cite{SA}. After the first and the second stages, the effective channel is estimated to be used in the digital precoding in the next stage. In the third stage, we use ZF digital precoder to mitigate the effect of inter-group interference. We then determine the power allocation policy to maximize the system sum rate, $R_{\text{sum}}$, while QoS constraints are met. For this purpose, we rewrite the optimization problem in \eqref{eq13} by taking into account the obtained user scheduling and antenna allocation strategy known as $c_{kbn}$ and $M_{kbn}$. This new optimization problem is still non-convex and difficult to be solved. The method used to solve this problem will be described in section \ref{subsec:power_Allo&ZF}.
\subsection{First stage: user scheduling}
\subsubsection{Problem formulation}
Assuming uniform power and equal antenna allocation as well as $\mathbf{G}_b=\mathbf{I},\forall b\in\mathcal{B}$, we define the following sub-problem to obtain the user scheduling strategy by
\begin{subequations}\label{eq14}
	\begin{align}
	\underset{\left\{\substack{c_{kbn},\\k\in\mathcal{K},b\in\mathcal{B},n\in\mathcal{N}}\right\}}{\max}&R_{\text{sum,1}}\\
	\text{s.t.}\quad\quad&\eqref{eq13c},\eqref{eq13d},\eqref{eq13k}.
	\end{align}
\end{subequations}

The exhaustive search method can find the optimal solution of the problem \eqref{eq14}, which for a fully loaded scenario, i.e., $K=2BN$ ($K_b=2N,\forall b\in\mathcal{B}$), the total number of states to be searched is $\left({K!}/{(2N)!^B}\right)\left({(2N)!}/{2^NN!}\right)^B$, where, the first and second terms are the total number of user assignment and grouping states, respectively. As a result, this method is inefficient due to its high computational complexity. In this regard, inspired by the idea presented in \cite{WCS}, we propose a heuristic algorithm called MWCS algorithm to assign users to mm-APs, and then we group the users according to the difference and correlation between their channels which is the estimated channels that contain only the LoS path (LoS CSI). 
\subsubsection{The proposed user assignment algorithm}\label{MWCS+Initialization}
The MWCS algorithm is based on the fact that the suboptimality of a user assignment strategy may be due to the imposition of a weak link to an MD or its suffering from the high inter-AP interference. Consequently, swapping the worst connection will probably give a stronger link to the MD or reduce inter-AP interference that can lead to a better individual data rate.

To describe how the MWCS algorithm works, we define $\Pi=\{\mathcal{K}_1,\dots,\mathcal{K}_B\}$ as a partition of $\mathcal{K}$, such that $\forall b\neq b^\prime,\mathcal{K}_b\cap\mathcal{K}_{b^\prime}=\emptyset$, and $\cup_{b=1}^{B}\mathcal{K}_b=\mathcal{K}$, where $\mathcal{K}_b$ denotes the set of the users assigned to mm-AP $b$. In fact, $\Pi$ represents the strategy of asssigning users to mm-APs. The MWCS algorithm starts with an initial $\Pi$, then users within each mm-AP are grouped by algorithm \ref{userGrouping_Algo}. Afterward, we can obtain the user scheduling strategy including user assignment and user grouping, i.e., $\mathbf{c}$, where $\mathbf{c}\in\{0,1\}^{KBN\times1}$ denotes the collection of $c_{kbn},\forall k\in\mathcal{K},\forall b\in\mathcal{B},\forall n\in\mathcal{N}$. Then, the individual data rate of all users can be calculated by \eqref{eq7}.
\begin{definition}
	The corresponding connection with the lowest individual data rate is defined as the worst connection, and the user associated with this connection is known as the worst user. 
\end{definition} 
\begin{definition}
	The swap operation for user $k$ in mm-AP $b$ and user $j$ in mm-AP $b^\prime$ such that $b\neq b^\prime$ is equivalent to
	\begin{align}
	&\mathcal{K}_b=(\mathcal{K}_b\setminus\{k\})\cup\{j\},\;\text{and}\;\mathcal{K}_{b^\prime}=(\mathcal{K}_{b^\prime}\setminus\{j\})\cup\{k\}.
	\end{align}
	If $|\mathcal{K}_{b^\prime}|<2N$, let us consider a hole (empty capacity) instead of user $j$, then the swap operation is equivalent to
	\begin{align}
	&\mathcal{K}_b=\mathcal{K}_b\setminus\{k\},\;\text{and}\;\mathcal{K}_{b^\prime}=\mathcal{K}_{b^\prime}\cup\{k\}.
	\end{align}
\end{definition}
The algorithm consists of two main steps.

First step: at $t$-th iteration to find a better user scheduling strategy, the worst user is swapped with the users and holes within other mm-APs, where for every swap such as $i$, a new partition such as $\Pi_i$ is obtained. Afterward, the users within the two mm-APs involved, $\overline{\mathcal{B}}$, must be re-grouped by Algorithm \ref{userGrouping_Algo}, which results in a new user scheduling strategy, $\mathbf{c}_i$. Among the obtained new strategies, $\mathbf{c}_i,\forall i$, the strategy that achieves the highest sum rate is selected. Subsequently, if the selected strategy, $\mathbf{c}_{i^\star}$, leads to a higher sum rate than the previous strategy, $\mathbf{c}_t$, the selected strategy $\mathbf{c}_{i^\star}$ will be replaced. This process repeats until no improvement in the sum rate is achieved. After this, the second step for further improvements is executed. 

Second step: the first step deadlock can be overcome by removing the worst current connection, $\mathcal{K}_t\setminus\{k_\text{worst}\}$, and redefining it from the remaining connections and returning to the first step.

The algorithm terminates when no connection is left in the second step to redefine the worst connection. Finding a good initial partition has a great impact on the convergence speed and optimality of the heuristic algorithms. Therefore, instead of randomly generating the initial partition, among the mm-APs, we assign each user to the one that has at least one vacancy and can provide the strongest LoS link. In order to identify the strongest LoS link, we define ${\mathbf{H}}_{kb,0}=\alpha_{kb0}\,\mathbf{a}_\text{MD}(\phi_{kb0})\mathbf{a}_\text{AP}^\text{H}(\theta_{kb0})$ as the LoS channel matrix between MD $k$ and mm-AP $b$, and $\mathbf{h}_{kb,0}=\mathbf{v}_k^\text{H}\mathbf{H}_{kb,0}\in\mathbb{C}^{1\times M_\text{AP}}$ as the LoS channel vector between mm-AP antennas $b$ and the RF chain of MD $k$. Accordingly, the mm-AP that can provide the strongest LoS link for MD $k$ as $b=\arg\underset{b\in\mathcal{B}}{\max}|\mathbf{h}_{kb,0}|^2$.
\begin{algorithm}
	\caption{MWCS (user assignment) Algorithm}
	\label{userScheduling_Algo}
	\DontPrintSemicolon
	{\small
		\textbf{Initialization:} initialize, the iteration index $t=1$, the initial partition $\Pi$ based on the description of section \ref{MWCS+Initialization}, the set $\overline{\mathcal{B}}=\mathcal{B}$ namely the mm-APs that need grouping, the initial user scheduling strategy $\mathbf{c}_t$ by Algorithm \ref{userGrouping_Algo}, and the auxiliary set $\mathcal{K}_t=\mathcal{K}$. Given $\mathbf{c}_t$, compute the sum rate and the individual data rate of the users, i.e., $R_{\text{sum},1}(\mathbf{c}_t)$ and $R_k(\mathbf{c}_t),\forall k\in\mathcal{K}$, respectively.\;
		\Repeat{$\mathcal{K}_t\neq\emptyset$}{
			$k_\text{worst}=\arg\underset{k\in\mathcal{K}_t}{\min}\,R_k(\mathbf{c}_t)$.\;
			Initialize the auxiliary index $i=0$.\;
			\textbf{Step 1:}\;
			\For{$\forall b\in\mathcal{B}\setminus\{b_{k_\text{worst}}\}$}{
				\For{$\forall k\in\{\mathcal{K}_b\cup \text{hole}\}$}{
					$i=i+1$.\;
					Swap user $k_\text{worst}$ with user (or hole) $k$ and create a new partition (new user assignment), $\Pi_i$.
					Update $\overline{\mathcal{B}}=\{b_{k_\text{worst}},b_k\}$.
					Given $\Pi_i$ and $\overline{\mathcal{B}}$, group users by Algorithm \ref{userGrouping_Algo} and then create a new user scheduling strategy, $\mathbf{c}_i$.}
			} 
			$i^\star=\arg\underset{i}{\max}R_{\text{sum},1}(\mathbf{c}_i)$.\;
			\eIf{$R_{\text{sum},1}(\mathbf{c}_t)< R_{\text{sum},1}(\mathbf{c}_{i^\star})$}{
				$\mathbf{c}_{t+1}=\mathbf{c}_{i^\star}$ and $\mathcal{K}_{t+1}=\mathcal{K}$.\;
			}{\textbf{Step 2:}\;$\mathcal{K}_{t+1}=\mathcal{K}_t\setminus\{k_\text{worst}\}$ and $\mathbf{c}_{t+1}=\mathbf{c}_t$.\;}
			$t=t+1$.
		}
		\KwOut{$\mathbf{c}_t$}
	}
\end{algorithm}
\subsubsection{The user grouping algorithm }
According to the NOMA technique, to achieve better performance, two users with a significant difference between their channel gains (the near and far users) must be in the same group. On the other hand, given the spacial directivity of the mmWave channel, the users whose channels are highly correlated must be in the same group to exploit multiplexing gain \cite{CorrelationGrouping2}. In other words, the users whose channels are uncorrelated should be assigned to different groups to reduce the interference. Consequently, to select a user pair to form a group, we select the one with the highest channel correlation and channel gain difference. Given the LoS channels of users $i$ and $j$, $\mathbf{h}_{ib,0}$ and $\mathbf{h}_{jb,0}$, The channel gain difference, $\text{Diff}(i,j)$, and channel correlation, $\text{Corr}(i,j)$, between them are calculated as in\cite{CorrelationGrouping1}.

Thus, we can formulate a multi-objective optimization problem with $\left[\text{Diff}(i,j),\text{Corr}(i,j)\right]$ as the vector of objective functions to select the desired user pair. The scalarization methods can be used to solve such optimization problems, one of which is the weighted sum method \cite{MultiObjective}. This method replaces the vector of objective functions with $w_1\overline{\text{Corr}}(i,j)+w_2\overline{\text{Diff}}(i,j)$, where $\overline{\text{Corr}}(i,j)$ and $\overline{\text{Diff}}(i,j)$ are the normalized channel correlation and channel gain difference by the min-max normalization method \cite{min-maxNormalization}. Consequently, the scalarized multi-objective optimization problem is as follows.
\begin{subequations}\label{eq17}
	\begin{align}
	\underset{(i,j)}{\max}\quad&w_1\overline{\text{Corr}}(i,j)+w_2\overline{\text{Diff}}(i,j)\\
	\text{s.t.}\quad&i<j,\forall i\in\overline{\mathcal{K}},\forall j\in\overline{\mathcal{K}},
	\end{align}
\end{subequations}
where $\overline{\mathcal{K}}$ denotes the collection of non-grouped users and $w_1,w_2\in(0,1)$ are the weights of $\overline{\text{Corr}}(i,j)$ and $\overline{\text{Diff}}(i,j)$, respectively, that must confirm $w_1+w_2=1$.

In Algorithm 1, after specifying $\Pi=\{\mathcal{K}_1,\dots,\mathcal{K}_B\}$ at the beginning of each iteration, the mm-APs whose users have been changed, i.e., $\overline{\mathcal{B}}$, need to be re-grouped. For any member of $\overline{\mathcal{B}}$ such as mm-AP $b$, if its number of users exceeds its number of RF chains, namely $|\mathcal{K}_b|>N$, we must form $|\mathcal{K}_b|-N$ two-user groups. Each time problem \eqref{eq17} is solved, a group is formed, and members of that group are exited from the set $\overline{\mathcal{K}}$, so if we repeat this process $|\mathcal{K}_b|-N$ times for mm-AP $b$ the grouping of users in this mm-AP ends. It should be noted that we use the exhaustive search method to solve problem \eqref{eq17}, which requires investigating all $\frac{|\overline{\mathcal{K}}|(|\overline{\mathcal{K}}|-1)}{2}$ possible states. The method of grouping users is summarized in Algorithm \ref{userGrouping_Algo}.
\begin{algorithm}
	\caption{User grouping Algorithm}
	\label{userGrouping_Algo}
	\DontPrintSemicolon
	{\small
		\textbf{Initialization:} Initialize the iteration index $t=1$. Specify the strategy of assigning users to mm-APs ($\Pi$ or $\Pi_i$) and set $\overline{\mathcal{B}}$ according to Algorithm \ref{userScheduling_Algo}. Sort the indexes of the users based on \eqref{eq4}.\;
		\For{$\forall b\in\overline{\mathcal{B}}$}{
			\eIf{$|\mathcal{K}_b|>N$}{
				Initialize the collection of non-grouped users $\overline{\mathcal{K}}=\mathcal{K}_b$.\;
				\For{$n=1$\KwTo$|\mathcal{K}_b|-N$}{Solve problem \eqref{eq17} and obtain $(i^\star,j^\star)$.
				Put users $i^\star$ and $j^\star$ in a group, $c_{i^\star bn}=c_{j^\star bn}=1$.
			    Update the collection of non-grouped users, $\overline{\mathcal{K}}=\overline{\mathcal{K}}\setminus\{i^\star,j^\star\}$.}}{Form $|\mathcal{K}_b|$ single-user groups.}}
		\KwOut{$\mathbf{c}$}}
\end{algorithm}
\subsubsection{Convergence}
The system sum rate is bounded due to the limited resources such as power. On the other hand, in Algorithm \ref{userScheduling_Algo}, we only accept changes that lead to a strict increase in the system sum rate. Thus, due to the limitation of the system sum rate, Algorithm \ref{userScheduling_Algo} converges after a few iterations.
\subsubsection{Complexity analysis}
Given the definition of floating-point operations (flops) in \cite{flops} and ignoring all terms except the dominant term and only considering non-zero entries of vectors and matrices \cite{convex}, we evaluate the complexity of the proposed algorithms by counting the number of flops. In the $t$-th iteration of Algorithm \ref{userScheduling_Algo} in a fully loaded scenario, i.e., $K=2BN$ ($|\mathcal{K}_b|=2N,\forall b\in\mathcal{B}$), there are $|\mathcal{K}_b|(B-1)$ swaps for the worst user, in addition, with each swap, the users within the two mm-APs involved must be re-grouped by Algorithm \ref{userGrouping_Algo}. By ignoring the ineffective terms versus the dominant term and given the complexity of sorting operation ($\arg \max \text{or} \arg \min$) \cite{sortingFlop}, the complexity of Algorithm \ref{userGrouping_Algo} is $\mathcal{O}\left(8NM_{\text{AP}}(M_{\text{MD}}+N)\right)$. Also, the complexity of computing the system sum rate is $\mathcal{O}\left(2KBN^2M_{\text{AP}}M_{\text{MD}}\right)$. Based on these complexities, among the different parts of Algorithm \ref{userScheduling_Algo}, line 12 has the highest complexity so that the complexity of the other parts can be ignored. Accordingly, the computational complexity of Algorithm \ref{userScheduling_Algo} is $\mathcal{O}\left(4KB^2N^3M_{\text{AP}}M_{\text{MD}}\right)$ per iteration. As a result, given the total number of iterations in the worst case, the computational complexity of the MWCS Algorithm at the worst case and for fully loaded scenario is given by $\mathcal{O}\left(\left({K!}/{(2N)!^B}\right)\left(4KB^2N^3M_{\text{AP}}M_{\text{MD}}\right)\right)$ which is much lower than the complexity of the exhaustive search method, i.e., $\mathcal{O}\left(({(2N)!}/{2^NN!})^B({K!}/{(2N)!^B})(2KBN^2M_{\text{AP}}\right.$\\$\left.M_{\text{MD}})\right)$. In particular, for $B=3, N=3$ and $K=2BN$, the MWCS Algorithm, at worst case, can reduce the number of flops by $99.4\%$ compared to the exhaustive search method.
\subsection{Second stage: antenna allocation}
\subsubsection{Problem formulation}
After obtaining the user scheduling strategy, $\mathbf{c}^\star$, we formulate the antenna allocation problem with the assumption of uniform power allocation between users and without using the digital precoder as follows.
\begin{subequations}\label{eq18}
	\begin{align}
	\underset{\mathbf{m}}{\max}\quad&R_\text{sum,2}\\
	\text{s.t.}\quad&m_q\in\{M_{\min},\dots,M_{\text{AP}}-M_{\min}\},\forall q\in\mathcal{Q},
	\end{align}
\end{subequations}
where $\mathcal{Q}$ is a set of $Q$ two-user NOMA groups across the system. We also define vector $\mathbf{m}\in\mathbb{N}^{1\times Q}$ for simplicity, where each element of vector $\mathbf{m}$, $m_q$, determines the number of antennas assigned to both users of the NOMA group associated with this element, i.e., $m_q$ and $M_{\text{AP}}-m_q$. The exhaustive search can be used to find the optimal solution of problem \eqref{eq18}, which at the fully loaded scenario, $Q=BN$, has a computational complexity equal to $\mathcal{O}\left((M_{\text{AP}}-2M_{\min})^{BN}(2KBN^2M_{\text{AP}}M_{\text{MD}})\right)$. However, due to its high computational complexity, it imposes a high overhead on the system. To address this problem, we use the SA algorithm, which has been proven to provide an optimal solution with sufficient iterations \cite{SAprove1,SAprove2}.
\subsubsection{The antenna allocation algorithm}
The SA algorithm is started with a feasible solution and then moves to next neighboring solution, $\mathbf{m}^\text{new}$, of the current solution, $\mathbf{m}$, to find a better solution. If the neighbor solution is not better than the current solution, SA algorithm chooses between moving to $\mathbf{m}^\text{new}$ or staying in $\mathbf{m}$ based on acceptance probability. The acceptance probability, $P=\exp{(-\Delta/T)}$, depends on the difference of the objective function for $\mathbf{m}^\text{new}$ and $\mathbf{m}$, $\Delta$, as well as $T$, which is a parameter called temperature. In the early steps, the temperature is set too high to consider other worst solutions. As the temperature gradually decreases in the final steps, the worse solutions are less likely to be accepted. Therefore, if the number of iterations is sufficiently high, the algorithm converges to the optimal solution. The proposed SA algorithm for solving the problem \eqref{eq18} is summarized in Algorithm \ref{Antenna_Algo}. To create the neighbor solution in Algorithm \ref{Antenna_Algo}, we randomly choose one of the numbers $1$ to $Q-1$ as the number of vector elements $\mathbf{m}$ that must be changed, such as $q^\prime\in\{1,\dots,Q-1\}$. Then, among the elements of $\mathbf{m}$, we randomly select $q^\prime$ elements and determine the new value of each element such as $m_q$ by randomly selecting from $\{M_{\min},\dots,M_{\text{AP}}-M_{\min}\}$.
\begin{algorithm}[h]
	\caption{Antenna Allocation Algorithm}
	\label{Antenna_Algo}
	\DontPrintSemicolon
	{\small
		\textbf{Initialization:} Set the temperature to $T=T_0$. Consider $\mathbf{m}=[\frac{M_\text{AP}}{2},\dots,\frac{M_\text{AP}}{2}]^{1\times Q}$ as the current solution. Compute the current sum rate for $\mathbf{m}$, $R_\text{sum,2}$. Moreover $\mathbf{m}^\star=\mathbf{m}$ and $R_\text{sum,2}^\star=R_\text{sum,2}$.\;
		\While{$T\geq\epsilon_1$}{
			\For{$t=1$ \KwTo $t_{\max}$}{
				Create a neighbor solution, $\mathbf{m}^\text{new}$ and Compute sum rate for $\mathbf{m}^\text{new}$, $R_\text{sum,2}^\text{new}$.
				$\Delta=(R_\text{sum,2}-R_\text{sum,2}^\text{new})/R_\text{sum,2}$ and $P=\exp(-\Delta/T)$.\;
				\eIf{$R_\text{sum,2}^\text{new}\geq R_\text{sum,2}$}{$R_\text{sum,2}=R_\text{sum,2}^\text{new}$ and $\mathbf{m}=\mathbf{m}^\text{new}$.}{With the probability of accepting $P$, Accept $R_\text{sum,2}^\text{new}$ and $\mathbf{m}^\text{new}$ as the current sum rate, $R_\text{sum,2}$, and the current solution, $\mathbf{m}$, respectively.\;}
				\If{$R_\text{sum,2}\geq R_\text{sum,2}^\star$}{$R_\text{sum,2}^\star=R_\text{sum,2}$ and $\mathbf{m}^\star=\mathbf{m}$.\;}
			}
			$T=\beta T$.\;
		}
		\KwOut{$\mathbf{m}^\star$}}
\end{algorithm}
\subsection{Third stage: digital precoder and power allocation}\label{subsec:power_Allo&ZF}
\subsubsection{ZF digital precoder}
We use a ZF digital precoder to reduce interference between the NOMA groups within each mm-AP. Since each NOMA group can contain two users, we perform singular value decomposition (SVD) on the equivalent channel for each NOMA group , i.e., $\widetilde{\mathbf{H}}_{bn}\in\mathbb{C}^{N\times |\mathcal{Q}_{bn}|},\forall b\in\mathcal{B}$ and $\forall n\in\mathcal{N}$, which denotes the equivalent channel matrix corresponding to the NOMA group served by RF chain $n$ in mm-AP $b$ and $\mathcal{Q}_{bn}$ is a set that includes all users of the NOMA group associated with RF chain $n$ in mm-AP $b$. Now taking SVD of $\widetilde{\mathbf{H}}_{bn}$ we have
\begin{equation}
\widetilde{\mathbf{H}}_{bn}^\text{H}=\mathbf{U}_{bn}\mathbf{\Sigma}_{bn}\mathbf{V}_{bn}^\text{H},
\end{equation}
where $\mathbf{U}_{bn}=[\mathbf{u}_{bn1},\dots,\mathbf{u}_{bn|\mathcal{Q}_{bn}|}]\in\mathbb{C}^{|\mathcal{Q}_{bn}|\times|\mathcal{Q}_{bn}|}$ is the left sigular matrix, $\mathbf{\Sigma}_{bn}$ is singular value matrix that its diagonal entries are known as sigular values of $\widetilde{\mathbf{H}}_{bn}^\text{H}$, and $\mathbf{V}_{bn}$ is the right sigular matrix. Thus, the equivalent channel vector of NOMA group associated with RF chain $n$ in mm-AP $b$ is given by
\begin{equation}
\widehat{\mathbf{h}}_{bn}=\widetilde{\mathbf{H}}_{bn}\mathbf{u}_{bn1}\in\mathbb{C}^{N\times1}.
\end{equation}

Note that, the equivalent channel vector for single-user NOMA group, $|\mathcal{Q}_{bn}|=1$, can be directly delivered with its effective channel vector, i.e., $\widehat{\mathbf{h}}_{bn}=\widetilde{\mathbf{h}}_{bn}$. Now, the equivalent channel matrix for all the NOMA groups on all the RF chains of mm-AP $b$ is given by
\begin{equation}
\widehat{\mathbf{H}}_b=[\widehat{\mathbf{h}}_{b1},\dots,\widehat{\mathbf{h}}_{bN}]\in\mathbb{C}^{N\times N},
\end{equation} 
where $\widehat{\mathbf{H}}_b=\mathbf{I}_N$ if no user is assigned to mm-AP $b$. Consequently, the ZF digital precoder is given by
\begin{equation}
\mathbf{G}_b=\widehat{\mathbf{H}}_b\left(\widehat{\mathbf{H}}_b^\text{H}\widehat{\mathbf{H}}_b\right)^{-1}\in\mathbb{C}^{N\times N}.
\end{equation}
\subsubsection{Power allocation design}
Given the effective channel matrix $\widetilde{\mathbf{H}}_b$ and the digital precoder $\mathbf{G}_b$, we formulate the power allocation problem in the following optimization problem as
\begin{subequations}\label{eq23}
	\begin{align}
	\underset{\left\{\substack{p_{kbn},\\k\in\mathcal{K},b\in\mathcal{B},n\in\mathcal{N}}\right\}}{\max}&R_\text{sum}\\
	\text{s.t.}\quad\quad&\eqref{eq13g},\eqref{eq13h},\eqref{eq13i},\eqref{eq13j}.
	\end{align}
\end{subequations}

In problem \eqref{eq23}, $c_{kbn}$ and $M_{kbn}$ are replaced by the obtained $c_{kbn}^\star$ and $M_{kbn}^\star$ in the first and the second stages. Problem \eqref{eq23} is a non-convex optimization problem, however, it can be equivalently converted to a canonical form of a DC programming technique \cite{DC} as follows
\begin{subequations}\label{eq24}
	\begin{align}
	&\underset{\mathbf{p}}{\min}\quad F_1(\mathbf{p})-F_2(\mathbf{p})\\
	\text{s.t.}\quad&\mathbf{0}\preceq\mathbf{p},\label{eq24a}\\
	&\mathbf{c^\star}^\text{T}\mathbf{p}\leq p_\text{total},\label{eq24b}\\
	&c_{kbn}^\star c_{ibn}^\star|\widetilde{\mathbf{h}}_{ib}^\text{H}\mathbf{g}_{bn}|^2D_{k\rightarrow i,2}^{bn}(\mathbf{p})\leq\notag\\
	&c_{kbn}^\star c_{ibn}^\star|\widetilde{\mathbf{h}}_{kb}^\text{H}\mathbf{g}_{bn}|^2D_{i\rightarrow i,2}^{bn}(\mathbf{p}),\forall k<i,\forall b\in\mathcal{B},\forall n\in\mathcal{N},\label{eq24c}\\
	&\left(2^{c_{kbn}^\star R_\text{min}}-1\right)D_{k\rightarrow k,2}^{bn}(\mathbf{p})\leq c_{kbn}^\star p_{kbn}|\widetilde{\mathbf{h}}_{kb}^\text{H}\mathbf{g}_{bn}|^2,\notag\\
	&\forall k\in\mathcal{K},\forall b\in\mathcal{B},\forall n\in\mathcal{N},\label{eq24d}
	\end{align}
\end{subequations}
where $\mathbf{p}\in\mathbb{R}^{KBN\times1}$ denotes the collection of $p_{kbn}$, also $F_1(\mathbf{p})$ and $F_2(\mathbf{p})$ are given by
\begin{align}
&F_1(\mathbf{p})=-\sum_{k\in\mathcal{K}}\sum_{b\in\mathcal{B}}\sum_{n\in\mathcal{N}}\log_2\left(D_{k\rightarrow k,1}^{bn}(\mathbf{p})\right),\,
\text{and}\notag\\
&F_2(\mathbf{p})=-\sum_{k\in\mathcal{K}}\sum_{b\in\mathcal{B}}\sum_{n\in\mathcal{N}}\log_2\left(D_{k\rightarrow k,2}^{bn}(\mathbf{p})\right),
\end{align}
respectively, in which
\begin{align}
&D_{k\rightarrow i,1}^{bn}(\mathbf{p})=|\mathbf{\widetilde{h}}_{kb}^\text{H}\mathbf{g}_{bn}|^2c_{ibn}p_{ibn}+\text{\rom{1}}_{k\rightarrow i}^{bn}+\text{\rom{2}}_{k}^{bn}+\text{\rom{3}}_{k}^{b}+\sigma^2,\label{eq26}\\
&\text{and}\notag\\
&D_{k\rightarrow i,2}^{bn}(\mathbf{p})=\text{\rom{1}}_{k\rightarrow i}^{bn}+\text{\rom{2}}_{k}^{bn}+\text{\rom{3}}_{k}^{b}+\sigma^2.\label{eq27}
\end{align}

Since the functions \eqref{eq26} and \eqref{eq27} are affine with respect to $\mathbf{p}$, therefore, $F_1(\mathbf{p})$ and $F_2(\mathbf{p})$ are differentiable convex functions with respect to $\mathbf{p}$. Consequently, according to the first-order condition for the convex functions \cite{convex} we have
\begin{align}
F_2(\mathbf{p})\geq\underbrace{ F_2(\mathbf{p}_t)+\nabla_{\mathbf{p}}F_2(\mathbf{p}_t)^\text{T}\left(\mathbf{p}-\mathbf{p}_t\right)}_{\triangleq \widehat{F}^{(t)}_2(\mathbf{p}) \,\text{as a auxiliary function}},
\end{align}
where $\nabla_{\mathbf{p}}F_2(\mathbf{p}_t)=\left\{\frac{\partial F_2(\mathbf{p})}{\partial p_{kbn}}|_{\mathbf{p}_t}\right\}_{k\in\mathcal{K},b\in\mathcal{B},n\in\mathcal{N}}\in\mathbb{R}^{KBN\times1}$ represent the gradient of $F_2(\mathbf{p})$ with respect to $\mathbf{p}$ and is given by
\begin{align}
\frac{\partial F_2(\mathbf{p})}{\partial p_{kbn}}|_{\mathbf{p}_t}=&
-\frac{1}{\ln(2)}\sum_{k^\prime=k+1}^{K}\frac{|\mathbf{\widetilde{h}}_{k^\prime b}^\text{H}\mathbf{g}_{bn}|^2c_{kbn}^\star}{D_{k^\prime\rightarrow k^\prime,2}^{bn}(\mathbf{p}_t)}\notag\\
&-\frac{1}{\ln(2)}\sum_{k^\prime\in\mathcal{K}}\sum_{n^\prime\in\mathcal{N},n^\prime\neq n}\frac{|\mathbf{\widetilde{h}}_{k^\prime b}^\text{H}\mathbf{g}_{bn}|^2c_{kbn}^\star}{D_{k^\prime\rightarrow k^\prime,2}^{bn^\prime}(\mathbf{p}_t)}\notag\\
&-\frac{1}{\ln(2)}\sum_{k^\prime\in\mathcal{K}}\sum_{b^\prime\in\mathcal{B},b^\prime\neq b}\sum_{n^\prime\in\mathcal{N}}\frac{|\mathbf{\widetilde{h}}_{k^\prime b}^\text{H}\mathbf{g}_{bn}|^2c_{kbn}^\star}{D_{k^\prime\rightarrow k^\prime,2}^{b^\prime n^\prime}(\mathbf{p}_t)}.
\end{align}

Now, we can obtain an upper bound for the minimization problem \eqref{eq24} by solving the following convex optimization problem
\begin{subequations}\label{eq30}
	\begin{align}
	\underset{\mathbf{p}}{\min}\quad&F_1(\mathbf{p})-\widehat{F}^{(t)}_2(\mathbf{p})\\
	\text{s.t.}\quad&\eqref{eq24a},\eqref{eq24b},\eqref{eq24c},\eqref{eq24d}.
	\end{align}
\end{subequations}
\begin{figure}[t!]
	\begin{center}
		\includegraphics[height=4.67cm,width=8.7cm]{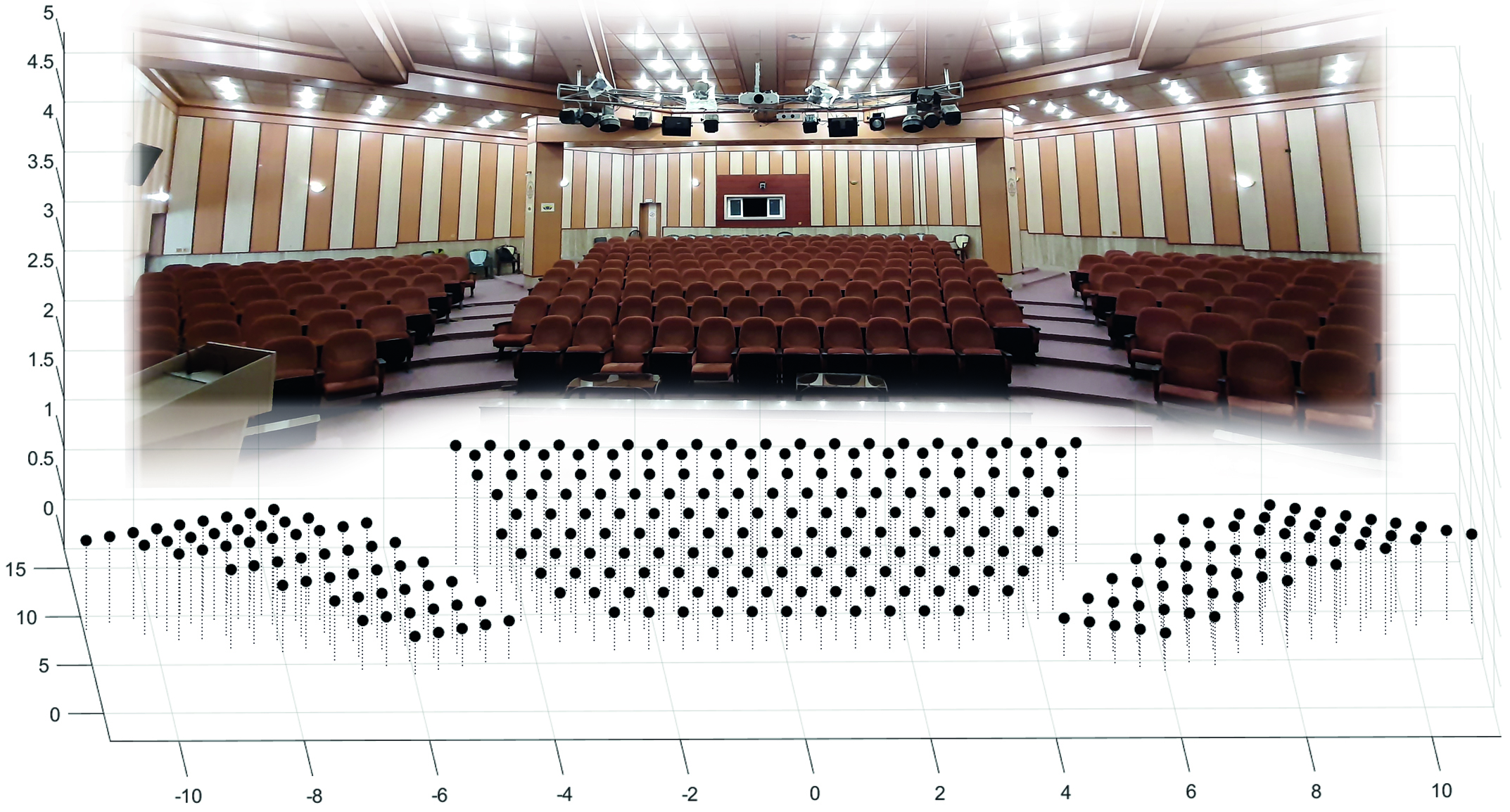}
	\end{center}
    \vspace{-0.7em}
	\caption{\footnotesize Shahid Fotouhi Hall of Isfahan University of Technology.}
	\label{fig4}
\end{figure}

To find a tight upper bound in \eqref{eq30}, we use the algorithm introduced by the DC programming method, \cite{DC,DC2}. Since the functions $F_1(\mathbf{p})$ and $F_2(\mathbf{p})$ are differentiable convex, the DC-based algorithm converges to a stationary point with a polynomial time computational complexity \cite{DC}.
\section{Simulation Results}\label{Sec:Sim_results}
\begin{figure}[t!]
	\begin{center}
		\includegraphics[height=3.64cm,width=4.7cm]{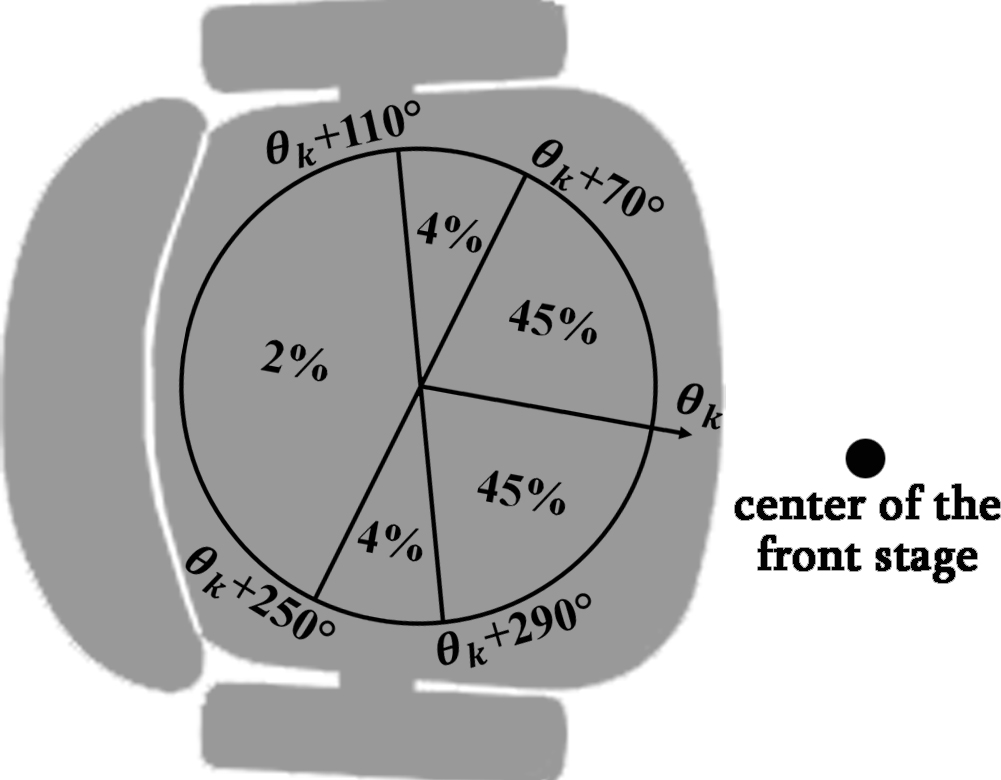}
	\end{center}
	\vspace{-0.7em}
	\caption{\footnotesize User's orientation probability distribution.}
	\label{fig5}
\end{figure}
To perform the simulation, we have selected Shahid Fotouhi Hall of Isfahan University of Technology, which is shown in Fig. \ref{fig4}. The coordinates of the seats and mm-APs for this scenario are available in \cite{seats}. Note that the locations of mm-APs are numerically optimized to reduce the average blockage probability for all seats \cite{3AccessPointDeploymentMmWave,1AccessPointDeploymentMmWave}. All $K$ users are randomly selected from the people sitting on the seats. Since most of the time, the users look toward the center of the front stage, for the user's orientation angle, $\widetilde{\Psi}_k$, we consider a probability distribution based on Fig. \ref{fig5}, in which $\theta_k$ is a location-dependent parameter that is calculated based on the coordinates of the user, $(X_k,Y_k)$, and the center of the front stage, $(0,0)$. We compare the simulation results of the proposed approach with the TDMA scheme, which is one of the mmWave-OMA conventional techniques. All statistical results are averaged over a large number of independent runs, i.e., 2000. In addition, the effect of all infeasible cases in the presented results is considered, which leads to the performance degradation. Also, similar to the assumption in \cite{multi-beamNOMAmmWave} and \cite{BeamwidthControl(EE)mmWave-NOMA}, we assume that $M_\text{MD}=15$. The simulation parameters are shown in Table \ref{table1}.
\begin{table}[h!]
	\caption{\small Simulation parameters}
	\label{table1}
	\scriptsize
	\centering
	\begin{tabular}{|c|c!{\vrule width 1pt}c|c|}
		\hline
		Parameter&Value&Parameter&Value\\
		\hline
		$f_c$&$60$ GHz&$H_\text{person}$&$125$ cm\\
		$L$&$2$&$H_\text{MD}$&$70$ cm\\
		$\gamma_0$&$2.25$&$r$&$27$ cm\\
		$\gamma_l$&$3.71$&$t_{\max}$&$12$\\
		$\sigma^2$&$-80$ dBm&$\beta$&$0.95$\\
		$M_\text{MD}$&$15$&$T_0$&$10$\\
		$M_{\min}$&${M_\text{AP}}/{6}$&$\epsilon_1$&$7\times10^{-11}$\\
		$R_{\min}$&$0.25$ bit/sec/Hz&$w_1$&$0.6$\\
		\hline
	\end{tabular}
\end{table}

\begin{figure}[t!]
	\begin{center}
		\includegraphics[height=6.52cm,width=8.7cm]{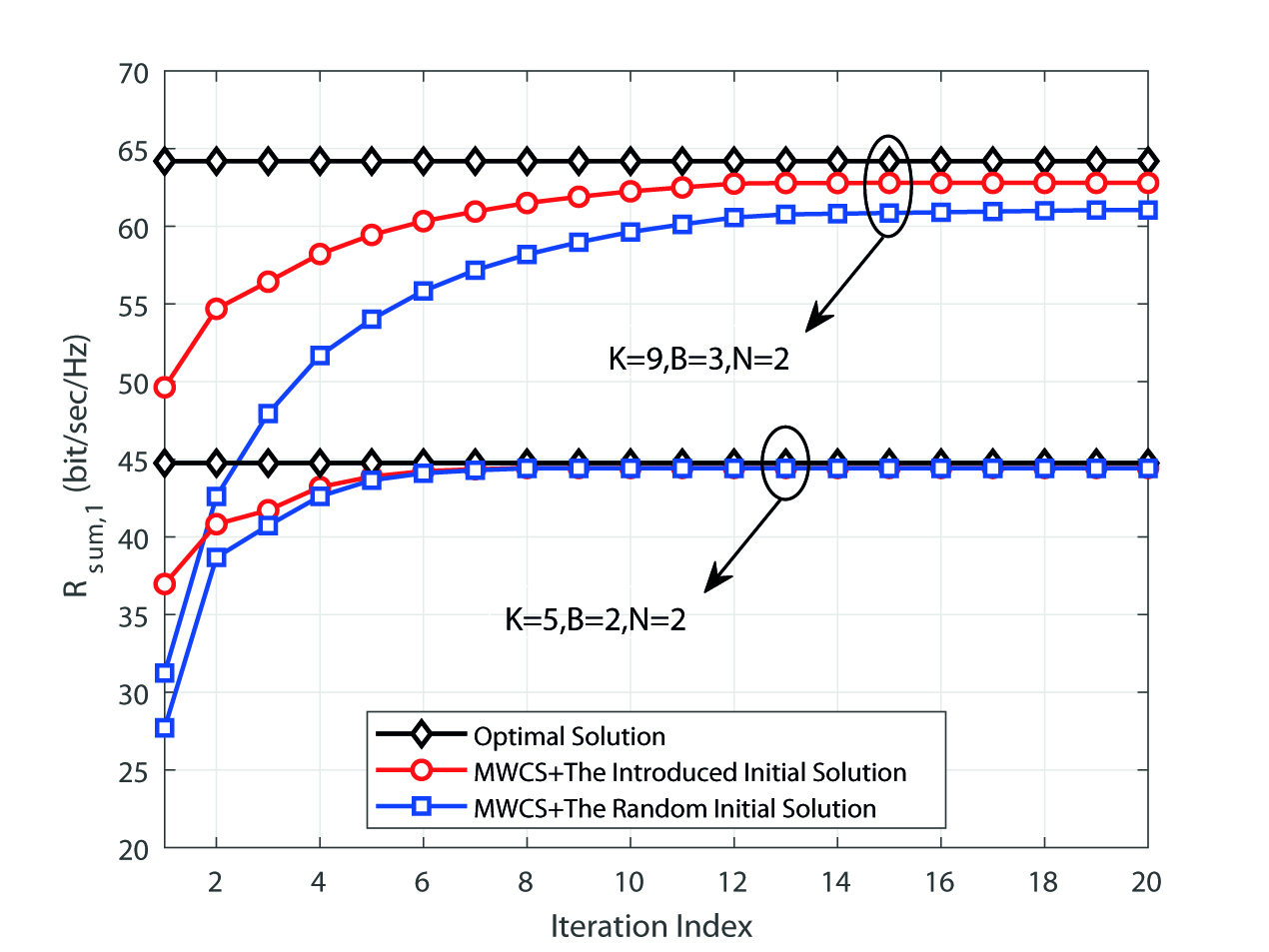}
	\end{center}
	\vspace{-0.7em}
	\captionsetup{singlelinecheck=false, justification=justified}
	\caption{\footnotesize Comparison of the MWCS algorithm and the optimal solution for $p_\text{total}=30$ dBm and $M_\text{AP}=120$.}
	\label{fig6}
\end{figure}

Fig. \ref{fig6} illustrates the average sum rate in the first stage, $R_{\text{sum},1}$, to compare the performance of the MWCS with the exhaustive search method. Given the high computational complexity of the exhaustive search method, we have simulated only two cases with a small number of users, mm-APs, and RF chains. Note that the proposed MWCS algorithm is applicable to the cases with more number of users, mm-APs, and RF chains. Fig. \ref{fig6} shows that the MWCS algorithm reaches  $97.7\%$ of the optimal value within only 13 iterations on average, which demonstrates the fast convergence and the effectiveness of our proposed algorithm; while the total number of iterations for the worst case is equal to the total number of states to assign users to mm-APs, i.e., $11130$ iterations. Thus, in practice, the total number of iterations of the MWCS algorithm is much less than the total number of iterations for the worst case. Fig. \ref{fig6} also illustrates that the performance of the MWCS algorithm when using the introduced initial solution is better than the random initial solution in terms of convergence speed and optimality.

\begin{figure}[t!]
	\begin{center}
		\includegraphics[height=6.52cm,width=8.7cm]{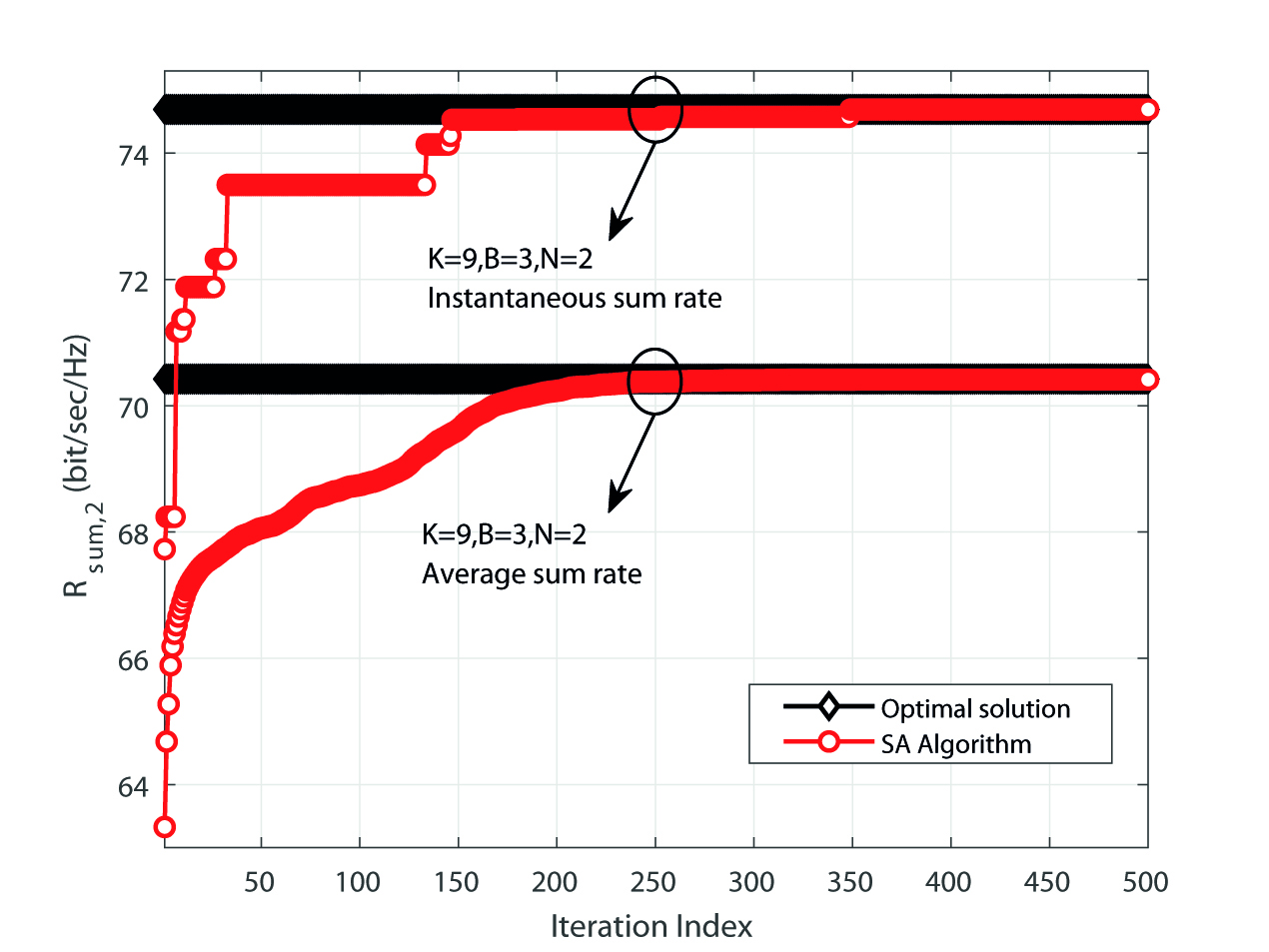}
	\end{center}
	\vspace{-0.7em}
	\captionsetup{singlelinecheck=false, justification=justified}
	\caption{\footnotesize Comparison of the SA algorithm and the optimal solution for $p_\text{total}=30$ dBm and $M_\text{AP}=120$.}
	\label{fig7}
\end{figure}

Fig. \ref{fig7} shows the convergence of the antenna allocation algorithm versus the exhaustive search method. As it can be seen, this algorithm achieves the globally optimal solution while its computational complexity is very lower than the exhaustive search method.

\begin{figure}[t!]
	\begin{center}
		\includegraphics[height=6.52cm,width=8.7cm]{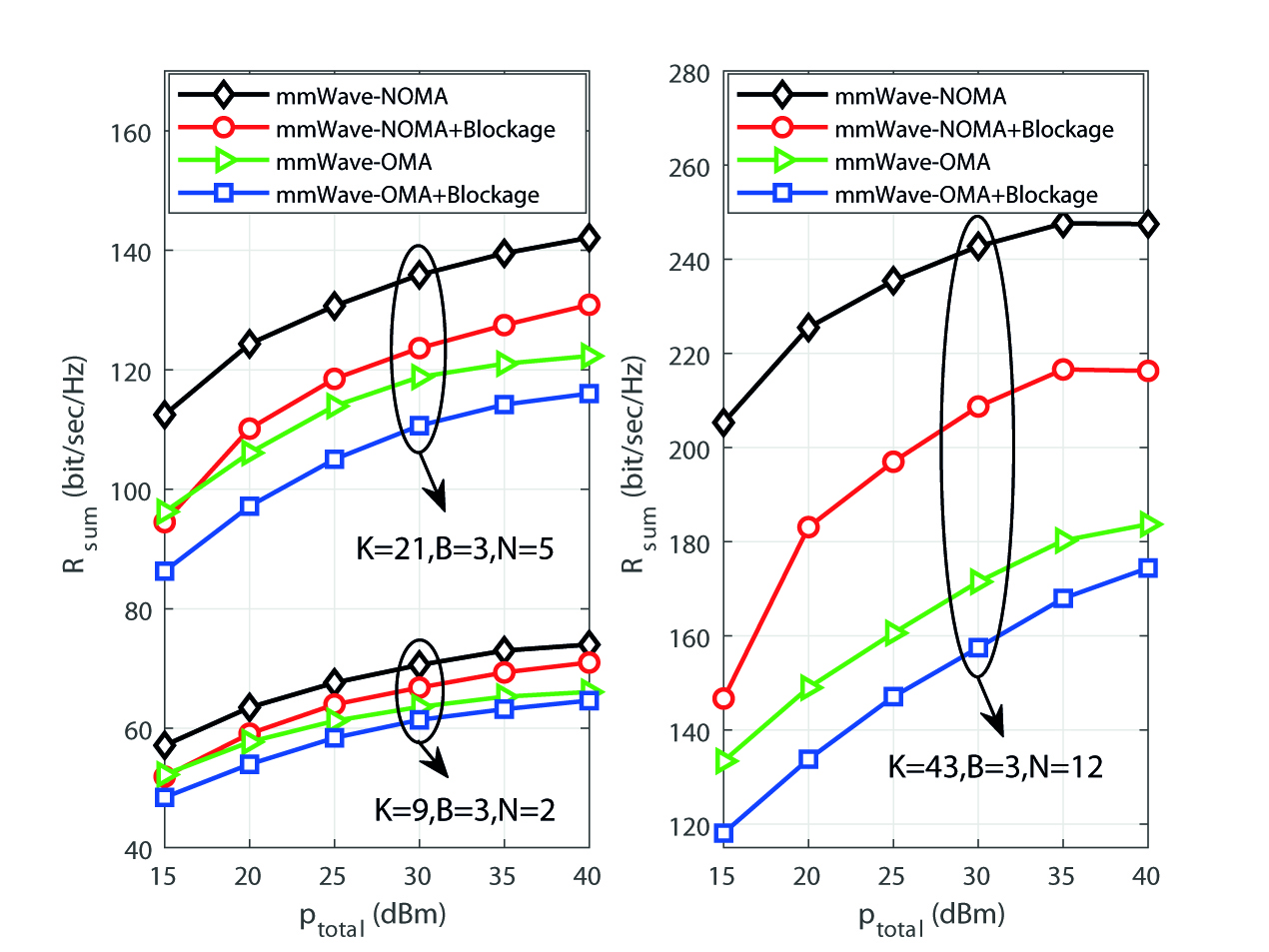}
	\end{center}
	\vspace{-0.7em}
	\caption{\footnotesize Average sum rate versus the total power budget when $M_{\text{AP}}=120$.}
	\label{fig8}
\end{figure}

Fig. \ref{fig8} illustrates the average sum rate in the third stage, $R_{\text{sum}}$, versus the total power budget. In this figure, in addition to the performance of the proposed system, the performance of the mmWave-OMA system is also shown as a performance benchmark for three cases with different values for $K$,$B$, and $N$. The simulation results in Fig. \ref{fig8} show that the proposed mmWave-NOMA system outperforms the mmWave-OMA system, and this performance improvement becomes more evident by increasing the number of surplus users over the number of RF chains. In particular, for $K=43$, $B=3$, and $N=12$, under blockage effect, the proposed mmWave-NOMA system performs on average $23\%$ better than the mmWave-OMA system, while this value is $30.5\%$ without considering the blockage effect. This is due to the fact that using NOMA, the intra-group interference in the user with better channel condition is controlled by SIC, and in the user with bad channel condition is negligible owing to applying the proposed resource allocation method. Besides, the inter-group interference decreases using the ZF precoder, and by appropriately assigning users to mm-APs, we can reduce inter-AP interference. Moreover, we observe that the average sum rate increases with the total power budget. However, due to the residual interference in the system, the sum rate uptrend is slowing. In this figure, the performance degradation due to the blockage effect is observable, which is not negligible for dense indoor scenarios. Additionally, the more RF chains, the more directions each mm-AP can send data, and the more users are covered simultaneously, resulting in the higher the sum rate. For a large number of users, $K=43$, and high total power budget, $p_\text{total}=40$ dBm, the amount of inter-AP interference is significant. Therefore, as shown in Fig. \ref{fig8}, the trend of increasing the average sum rate stops.

\begin{figure}[t!]
	\begin{center}
		\includegraphics[height=6.52cm,width=8.7cm]{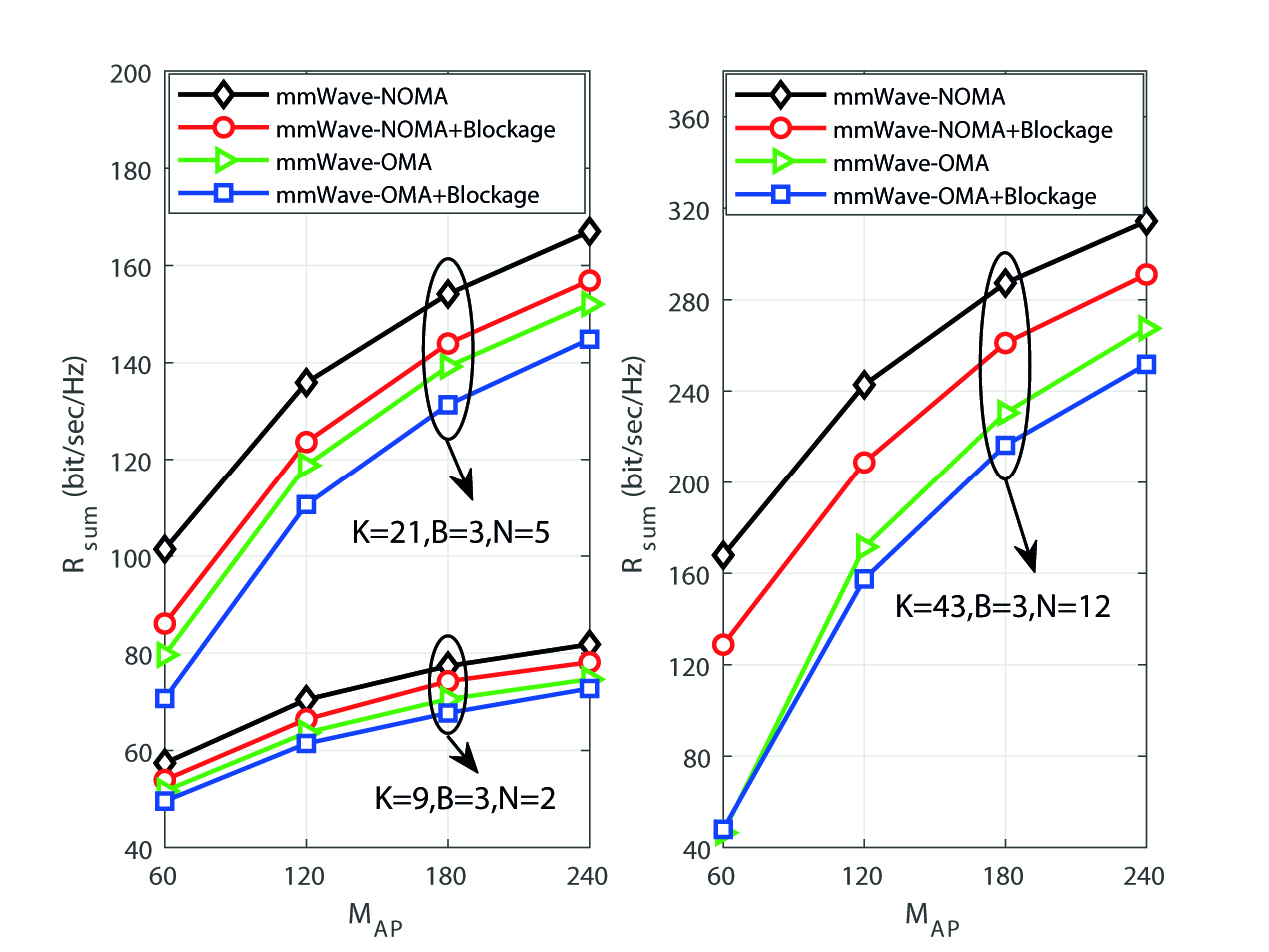}
	\end{center}
	\vspace{-0.7em}
	\captionsetup{singlelinecheck=false, justification=justified}
	\caption{\footnotesize Average sum rate versus the number of antennas at each mm-AP for $p_\text{total}=30$ dBm.}
	\label{fig9}
\end{figure}

Fig. \ref{fig9} shows the average sum rate in the third stage, $R_{\text{sum}}$, versus the number of antennas equipped at each mm-AP, $M_{\text{AP}}$. As the number of antennas per mm-AP increases, the array gain goes up, and beams with lower sidelobes are formed. Thus, the inter-AP interference decreases, and the average sum rate monotonically increases with the number of antennas at each mm-AP. Fig. \ref{fig9} also confirms that under the blockage effect and without blockage effect, for $K=43$, $B=3$, $N=12$, the proposed mmWave-NOMA system performs on average $29.5\%$ and $34\%$ better than the mmWave-OMA system, respectively. With the increasing number of users, a higher number of them may expose to the blockage; thus, considering this blockage effect, the more number of users leads to a higher decrease in the system sum rate. Moreover, when the number of users is high, $K=43$, and the number of antenna elements is small, $M_\text{AP}=60$, the infeasible cases increase, especially in the mmWave-OMA scheme. Therefore the average sum rate decreases significantly. Also, note that Figures \ref{fig8} and \ref{fig9} show that the mmWave-NOMA system under blockage even performs $9.2\%$ better than the corresponding OMA system without blockage.
\begin{figure}[t!]
	\begin{center}
		\includegraphics[height=6.52cm,width=8.7cm]{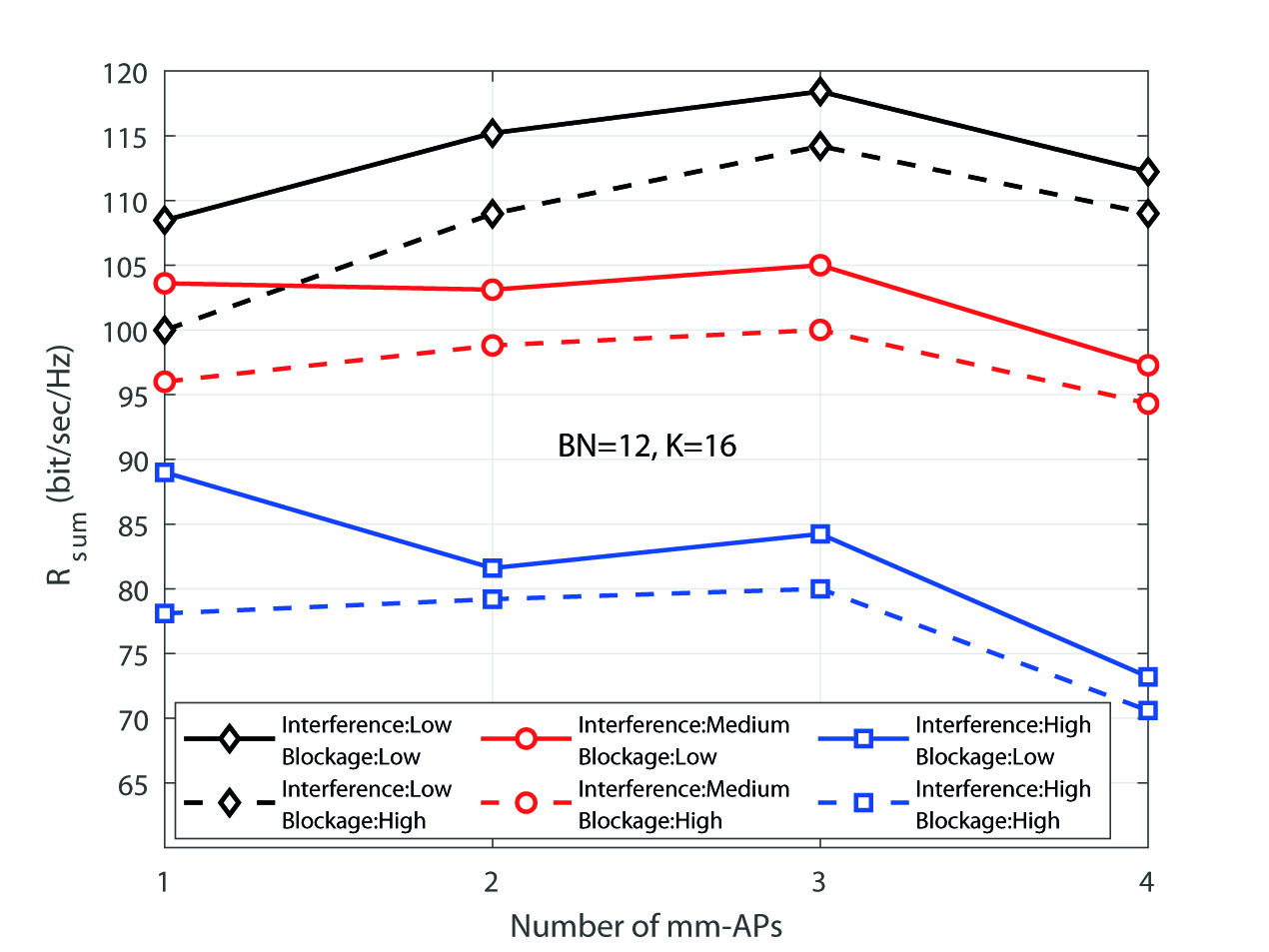}
	\end{center}
	\vspace{-0.7em}
	\captionsetup{singlelinecheck=false, justification=justified}
	\caption{\footnotesize Average sum rate versus the number of mm-APs for low, medium, and high interference setups i.e., ($p_\text{total}=20$ dBm, $M_\text{AP}=240$), ($p_\text{total}=25$ dBm, $M_\text{AP}=120$), and ($p_\text{total}=30$ dBm, $M_\text{AP}=60$), respectively.}
	\label{fig10}
\end{figure}

Fig. \ref{fig10} shows the average sum rate in the third stage, $R_{\text{sum}}$, versus the number of mm-APs for the proposed mmWave-NOMA system. We have simulated three different setups for two blockage scenarios consisting of a low blockage scenario (MD in hand) with $H_\text{MD}=70$ cm and a high blockage scenario (MD in pocket) with $H_\text{MD}=50$ cm. As it can be seen in Fig. \ref{fig10}, for all setups and scenarios, the compromise is made at $B=3$, while for the high interference setup in low blockage conditions, $B=1$ is the best. In fact, for the high interference setup in low blockage conditions, the inter-AP interference has a predominant effect compared to the blockage probability. Consequently, by increasing the number of mm-APs, the system sum rate decreases. Consequently, the number of required mm-APs needs to be chosen based on a tradeoff between the blockage probability and co-channel interference.

\begin{figure}[t!]
	\begin{center}
		\includegraphics[height=6.52cm,width=8.7cm]{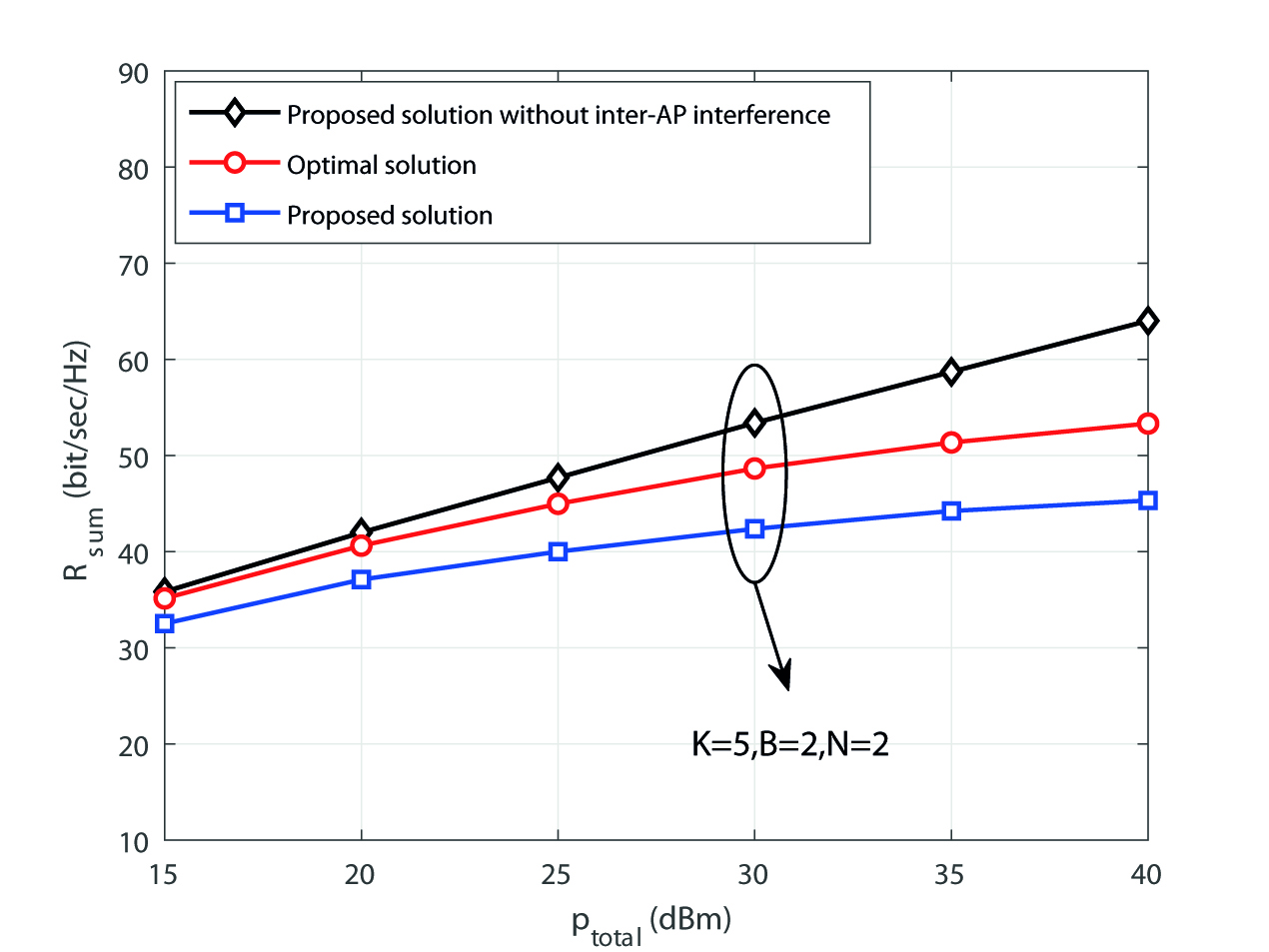}
	\end{center}
	\vspace{-0.7em}
	\captionsetup{singlelinecheck=false, justification=justified}
	\caption{\footnotesize Performance evaluation of the proposed solution for $M_{\text{AP}}=60$.}
	\label{fig11}
\end{figure}
Fig. \ref{fig11} shows the average sum rate for three different approaches, i.e., the exhaustive search method on all possible states and the proposed solution with and without inter-AP interference. As shown in Fig. \ref{fig11}, for this particular case namely $K=5,B=2,$ and $N=2$, the performance of our proposed method is on average $11.4\%$ lower than the exhaustive search method, while reducing the computational complexity by about $96\%$. It also shows that unlike outdoor mmWave networks, which ignore inter-AP interference due to its small amount, this interference in our network cannot be ignored due to the very close deployment of mm-APs. It should be noted that an alternative optimization method, i.e., repeating all three steps until achieving convergence, can be used to improve performance; however, the performance improvement is negligible and complexity is high.
\section{Conclusions}\label{Sec:Conclusion}
In this paper, we have proposed a suitable framework for resource allocation in the downlink of mmWave-NOMA communication through multi-AP for dense venues. The resource allocation in this network is very challenging due to the high co-channel interference caused by dense deployment and the inherent complexity of the system stemming from the combination of multi-AP structure, mmWave communication, and NOMA technique. To solve the highly computational complex resource allocation problem of the desired network, we have broken the main problem into three sub-problems. Then, we have proposed an algorithm for each sub-problem and evaluated the performance of each of them in terms of optimality, complexity, and convergence. Simulation results show that the sum rate of the proposed mmWave-NOMA scheme under the blockage effect is still $9.2\%$ higher than the corresponding OMA scheme without blockage. Also, we have found that a tradeoff between the amount of co-channel interference and the availability of LoS mmWave links should be considered to find the required number of AP in a mmWave-NOMA scheme.

\bibliographystyle{IEEEtran}
\bibliography{ref}

\end{document}